\documentclass{emulateapj}
\shorttitle{Finding dark stars with the JWST}
\shortauthors{Zackrisson et al.}

\begin{document}

\title{Finding high-redshift dark stars with the James Webb Space Telescope}

\author{Erik Zackrisson\altaffilmark{1,8}$^*$, Pat Scott\altaffilmark{2,8}, Claes-Erik Rydberg\altaffilmark{1,8}, Fabio Iocco\altaffilmark{3}, Bengt Edvardsson\altaffilmark{4}, G\"oran \"Ostlin\altaffilmark{1,8}, Sofia Sivertsson\altaffilmark{5,8}, Adi Zitrin\altaffilmark{6}, Tom Broadhurst\altaffilmark{6} \& Paolo Gondolo\altaffilmark{7}}
\altaffiltext{*}{E-mail: ez@astro.su.se}
\altaffiltext{1}{Department of Astronomy, Stockholm University, 10691 Stockholm, Sweden}
\altaffiltext{2}{Department of Physics, Stockholm University, 10691 Stockholm, Sweden}
\altaffiltext{3}{Institut d'Astrophysique de Paris, UMR 7095-CNRS Paris, Universit\'e Pierre et Marie Curie, Boulevard Arago 98bis, 75014, Paris, France}
\altaffiltext{4}{Department of Physics and Astronomy, Uppsala Astonomical Observatory, Box 516, 751 20 Uppsala, Sweden}
\altaffiltext{5}{Department of Theoretical Physics, Royal Institute of Technology (KTH), 10691 Stockholm, Sweden}
\altaffiltext{6}{School of Physics and Astronomy, Tel Aviv University, Israel}
\altaffiltext{7}{Physics Department, University of Utah, Salt Lake City, UT84112, USA}
\altaffiltext{8}{Oskar Klein Centre for Cosmoparticle Physics, AlbaNova University Centre, 10691 Stockholm, Sweden}

\begin{abstract}
The first stars in the history of the Universe are likely to form in the dense central regions of $\sim 10^5$--$10^6\ M_\odot$ cold dark matter halos at $z\approx 10$--50. The annihilation of dark matter particles in these environments may lead to the formation of so-called dark stars, which are predicted to be cooler, larger, more massive and potentially more long-lived than conventional population III stars. Here, we investigate the prospects of detecting high-redshift dark stars with the upcoming James Webb Space Telescope (JWST). We find that all dark stars with masses up to $10^3\ M_\odot$ are intrinsically too faint to be detected by JWST at $z>6$. However, by exploiting foreground galaxy clusters as gravitational telescopes, certain varieties of cool ($T_\mathrm{eff}\leq 30000$ K) dark stars should be within reach at redshifts up to $z\approx 10$. If the lifetimes of dark stars are sufficiently long, many such objects may also congregate inside the first galaxies. We demonstrate that this could give rise to peculiar features in the integrated spectra of galaxies at high redshifts, provided that dark stars make up at least $\sim 1\%$ of the total stellar mass in such objects. 
\end{abstract}

%% Keywords should appear after the \end{abstract} command. The uncommented
%% example has been keyed in ApJ style. See the instructions to authors
%% for the journal to which you are submitting your paper to determine
%% what keyword punctuation is appropriate.

%% Authors who wish to have the most important objects in their paper
%% linked in the electronic edition to a data center may do so in the
%% subject header.  Objects should be in the appropriate "individual"
%% headers (e.g. quasars: individual, stars: individual, etc.) with the
%% additional provision that the total number of headers, including each
%% individual object, not exceed six.  The \objectname{} macro, and its
%% alias \object{}, is used to mark each object.  The macro takes the object
%% name as its primary argument.  This name will appear in the paper
%% and serve as the link's anchor in the electronic edition if the name
%% is recognized by the data centers.  The macro also takes an optional
%% argument in parentheses in cases where the data center identification
%% differs from what is to be printed in the paper.

\keywords{Dark ages, reionization, first stars -- dark matter -- galaxies: high-redshift -- stars: Population III}

%% From the front matter, we move on to the body of the paper.
%% In the first two sections, notice the use of the natbib \citep
%% and \citet commands to identify citations.  The citations are
%% tied to the reference list via symbolic KEYs. The KEY corresponds
%% to the KEY in the \bibitem in the reference list below. We have
%% chosen the first three characters of the first author's name plus
%% the last two numeral of the year of publication as our KEY for
%% each reference.

\section{Introduction}
\label{intro}
The first stars in the history of the Universe are predicted to form inside $\sim$10$^5$--$10^6 M_\odot$ minihalos at redshifts $z\approx 10$--50 \citep[e.g.][]{Tegmark et al.,Yoshida et al.}. Due to the lack of efficient coolants in the primordial gas at these early epochs, the resulting population III stars are believed to be very massive \citep[$\gtrsim 100\ M_\odot$; e.g.][]{Abel et al.,Bromm et al. b}, hot \citep[effective temperature $T_\mathrm{eff}\sim 10^5$ K; e.g][]{Bromm et al. a} and short-lived \citep[$\approx 2$--3 Myr;][]{Schaerer a}. Non-rotating population III stars with masses of 50--140 $M_\odot$ or $M>260 \ M_\odot$ are expected to collapse directly to black holes, whereas stars with masses of 140--$260 \ M_\odot$ may produce luminous pair-instability supernovae \citep[e.g.][]{Heger et al.}. The latter may enrich the ambient medium with heavy elements and initiate the transition to the normal mode of star formation (population I and II, with a characteristic stellar mass $<1 \ M_\odot$) known from the low-redshift Universe. The highly energetic radiation emitted from population III stars during their lifetimes may also have played an important role in cosmic reionization at $z>6$ \citep[e.g.][]{Sokasian et al.,Trenti & Stiavelli}. An observational confirmation of very massive population III stars would be an important breakthrough in the study of the star formation, chemical enrichment and reionization history of the Universe.

The James Webb Space Telescope\footnote{http://www.jwst.nasa.gov}(JWST), scheduled for launch in 2014, has been designed to study the epoch of the first light, reionization and galaxy assembly, but is not expected to be able to directly detect individual population III stars at $z\gtrsim 10$. Searches for population III stars with the JWST would instead focus on the pair-instability supernovae produced at the end of their lifetimes \citep{Weinmann & Lilly}, or on $\sim$10$^5$--$10^7\ M_\odot$ clusters of population III stars \citep{Bromm et al. a,Scannapieco et al.,Trenti et al.,Johnson et al. b,Johnson}.  Other alternatives are to look for the spectral signatures of population III stars forming in pockets of unenriched gas within high-redshift galaxies \citep{Tumlinson & Shull,Schaerer a,Schaerer b,Dijkstra & Wyithe,Johnson et al. a}, or their integrated contribution to the infrared extragalactic background light \citep{Santos et al.,Cooray et al.}.

It has recently been recognized that annihilation of dark matter in the form of Weakly Interacting Massive Particles (WIMPs; e.g.~the lightest supersymmetric or Kaluza-Klein particles, or an extra inert Higgs boson) may have generated a first population of stars with properties very different from the canonical population III \citep{Spolyar et al. a}.  Because the first stars are likely to form in the high-density central regions of minihalos, annihilation of dark matter into standard model particles could serve as an additional energy source alongside or instead of nuclear fusion within these objects. This leads to the formation of so-called dark stars, which are predicted to be cooler, larger, more massive and potentially longer-lived than conventional population III stars \citep{Spolyar et al. a,Iocco b,Freese et al. a,Iocco et al.,Yoon et al.,Taoso et al.,Natarajan et al.,Freese et al. b,Spolyar et al. b,Umeda et al.,Ripamonti et al. a}. Similar effects have been seen in studies of the impacts of dark matter upon population I and II stars \citep{Salati & Silk,Fairbairn et al.,Scott et al. a,Scott et al. b,Casanellas & Lopes}. 

A significant population of high-redshift dark stars could have important consequences for the formation of intermediate and supermassive black holes \citep{Spolyar et al. b}, for the cosmic evolution of the pair-instability supernova rate \citep{Iocco c}, for the X-ray extragalactic background and for the reionization history of the Universe \citep{Schleicher et al.}. Effects such as these can be used to indirectly constrain the properties of dark stars, but no compelling evidence for or against a dark star population at high redshifts has so far emerged. Here, we explore a more direct approach -- the prospects for detection of population III dark stars using the JWST.

When attempting to assess the detectability of dark stars at high redshifts, the expected lifespan of such objects represents a crucial aspect. In principle, dark stars could live indefinitely, provided that there is ample dark matter available to fuel them. Dark stars are powered by gravitationally-contracted dark matter, pulled into their core as infalling gas steepens the gravitational potential during the formation phase. Because annihilation depletes the dark matter present within a star, ordinary fusion processes eventually take over as the dominant power source if the dark matter is not replenished. At this point, the dark star will essentially transform into a conventional population III star, albeit more massive because the increased duration of the formation phase has allowed it to accrete more gas. The dark matter present within the star during formation will last only for a few million years, but these reserves may later be replenished by scattering of WIMPs on nucleons, causing them to lose energy and become captured in the stellar core.  This could boost the longevity of dark stars substantially \citep{Iocco et al.,Freese et al. a-1,Yoon et al.,Taoso et al.,Spolyar et al. b,Iocco c}. The ongoing replenishment of the dark matter through capture and the resultant increase in longevity rely upon a number of strong assumptions and approximations, and the feasibility of such a mechanism is still yet to be proven by detailed calculations \citep{Sivertsson & Gondolo}.

Whether the capture process will be efficient depends on two factors: the scattering cross-sections of the dark matter particles with ordinary nucleons, and the amount and density of dark matter available for capture from the star's surroundings. The WIMP-nucleon scattering cross-sections can be constrained by direct detection experiments \citep[e.g.][]{Savage et al.,CDMS08} and searches for neutrinos produced by annihilation in the Sun \citep[e.g.][]{IceCube09}. However, the question of the amount of dark matter available for refueling is more complicated \citep{Sivertsson & Gondolo}. In a pristine halo, WIMP annihilations and scatterings during the formation stage would eventually deplete orbits with low angular momenta and result in a cavity of reduced dark matter density in the vicinity of the dark star. Whether infalling WIMPs could restore the balance before the star evolves into a supernova or a black hole may depend on the overall structural evolution of the minihalo (e.g. contraction, mergers with other halos), on tidal interactions of the dark star with subhalos, gas clouds and possibly other population III stars within the minihalo itself. Should a violent event cause the dark star to venture far from the centre of the minihalo, the dark matter density would very quickly become too low to sustain further dark matter burning. At the current time, estimates of the dark star lifetime in the presence of capture range from a few times $10^5$ to $10^{10}$ years \citep[e.g.][]{Yoon et al.,Iocco c,Sivertsson & Gondolo}. Many of the mechanisms mentioned above for replenishing the dark matter in the centre of the dark star have moreover not yet been explored. In this paper, we will therefore treat the duration of the dark star phase as a free parameter.

Model atmospheres and evolutionary histories of dark stars are presented in Sect.~\ref{models}. The detectability of isolated dark stars, with and without the effects of gravitational lensing by foreground galaxy clusters, is explored in Sect.~\ref{detectability}. In Sect.~\ref{discussion}, we explain how high-redshift dark stars can be distinguished from other objects based on their JWST colours, and discuss the possibility of detecting the spectral signatures of dark stars in the first generations of galaxies. Sect.~\ref{summary} summarizes our findings. Throughout this paper, we will assume a $\Lambda$CDM cosmology with $\Omega_\Lambda=0.73$, $\Omega_M=0.27$ and $H_0=72$ km s$^{-1}$ Mpc$^{-1}$.

After this paper had been submitted, another paper \citep{Freese et al. c} dealing with the prospects of detecting high-redshift dark stars with the JWST was posted on the arXiv preprint server. The two studies differ in their assumptions concerning the masses of dark stars. Whereas we consider objects with masses up to $\sim 10^3\ M_\odot$, Freese et al. instead focus on the detectability of `supermassive dark stars' (with masses up to $10^7\ M_\odot$). 

\begin{deluxetable*}{llllllllll}
\tabletypesize{\scriptsize}
%\rotate
\centering
\tablecaption{Dark star models\label{modeltable}}
\tablewidth{0pt}
\tablehead{
\colhead{WIMP mass} & \colhead{$M$ (M$_\odot$)\tablenotemark{a}} & \colhead{$R$ (cm)\tablenotemark{a}} & \colhead{$\log_{10}(g)$\tablenotemark{b}} & \colhead{$T_\mathrm{eff}$ (K)\tablenotemark{a}} & \colhead{$t_\mathrm{burn}$ (yr)} & \colhead{$t_\mathrm{SA}$ (yr)} & \colhead{$t_\mathrm{max}$ (yr)} & \colhead{Atmosphere}  & \colhead{$\max(z_\mathrm{obs})$\tablenotemark{c}}
}
\startdata
1\, GeV                 &106       & $2.4 \times 10^{14}$     &         $-$0.612       & $5.4 \times 10^3$   & $>10^{10}$ & $2.0 \times 10^{5}$ & $2.0 \times 10^{7}$  & \textsc{marcs}  & 10 \\
                       &371       & $2.7 \times 10^{14}$     &         $-$0.170       & $5.9 \times 10^3$   & $>10^{10}$ & $1.2\times 10^{6}$ & $1.2\times 10^{8}$ & \textsc{marcs}  & 11 \\
                       &690       & $1.1 \times 10^{14}$     &\phantom{$-$}0.879      & $7.5 \times 10^3$   & $>10^{10}$ &$2.4\times 10^{7}$  & $5.0\times10^8$ & \textsc{marcs}  & 11 \\
                       &756       & $3.7 \times 10^{13}$     &\phantom{$-$}1.865      & $1.0 \times 10^4$   & $>10^{10}$ & $2.3 \times 10^{8}$ & $5.0\times10^8$ & \textsc{tlusty} & 9 \\
                       &793       & $5.7 \times 10^{12}$     &\phantom{$-$}3.511      & $3.0 \times 10^4$   & $>10^{10}$ & $8.7\times 10^{8}$ & $5.0\times10^8$ & \textsc{tlusty} & 11\\
                       &824       & $5.8 \times 10^{11}$     &\phantom{$-$}5.512      & $1.1 \times 10^5$   & $6.2 \times 10^{6}$  & $4.9\times 10^{9}$ &  $6.2 \times 10^{6}$  & \textsc{tlusty} & 0.5\\
100\,GeV               &106       & $7.0 \times 10^{13}$     &\phantom{$-$}0.458      & $5.8 \times 10^3$   & $>10^{10}$ &  $6.0\times 10^{6}$ & $5.0\times10^8$ & \textsc{marcs}  & 6\\
                       &479       & $8.4 \times 10^{13}$     &\phantom{$-$}0.955      & $7.8 \times 10^3$   & $>10^{10}$ & $2.2\times 10^{7}$  & $5.0\times10^8$ & \textsc{marcs}  & 11\\
                       &716       & $1.1 \times 10^{13}$     &\phantom{$-$}2.895      & $2.3 \times 10^4$   & $>10^{10}$ & $2.9\times 10^{8}$ & $5.0\times10^8$ & \textsc{tlusty} & 13\\
                       &756       & $2.0 \times 10^{12}$     &\phantom{$-$}4.399      & $5.5 \times 10^4$   & $2.2 \times 10^{9}$  & $1.6\times 10^{9}$  & $5.0\times10^8$ & \textsc{tlusty} & 3\\
                       &787       & $5.8 \times 10^{11}$     &\phantom{$-$}5.492      & $1.1 \times 10^5$   & $5.9 \times 10^{6}$  & $4.5\times 10^{9}$ & $5.9 \times 10^{6}$   & \textsc{tlusty} & 0.5\\
10\,TeV                &106       & $2.2 \times 10^{13}$     &\phantom{$-$}1.463      & $6.0 \times 10^3$   & $>10^{10}$ & $1.7\times 10^{8}$  & $5.0\times10^8$  & \textsc{marcs}  & 2\\
											 &256       & $2.2 \times 10^{13}$     &\phantom{$-$}1.846      & $8.0 \times 10^3$   & $>10^{10}$ & $3.1\times 10^{8}$ & $5.0\times10^8$  & \textsc{marcs}  & 4\\
                       &327       & $2.0 \times 10^{13}$     &\phantom{$-$}2.036      & $1.0 \times 10^4$   & $>10^{10}$ & $2.8\times 10^{8}$ & $5.0\times10^8$  & \textsc{tlusty} & 5\\
                       &399       & $6.6 \times 10^{12}$     &\phantom{$-$}3.085      & $2.5 \times 10^4$   & $>10^{10}$ & $2.9\times 10^{8}$ & $5.0\times10^8$  & \textsc{tlusty} & 7\\
                       &479       & $2.9 \times 10^{12}$     &\phantom{$-$}3.879      & $3.2 \times 10^4$   & $>10^{10}$ & $1.9\times 10^{9}$  & $5.0\times10^8$  & \textsc{tlusty} & 2\\
                       &550       & $6.0 \times 10^{11}$     &\phantom{$-$}5.307      & $9.5 \times 10^4$   & $1.3 \times 10^{7}$  & $3.6\times 10^{9}$ & $1.3 \times 10^{7}$ & \textsc{tlusty} & 0.5\\
                       &553       & $4.8 \times 10^{11}$     &\phantom{$-$}5.503      & $1.1 \times 10^5$   & $5.1 \times 10^{6}$  & $3.9\times 10^{9}$ & $5.1 \times 10^{6}$ & \textsc{tlusty} & $<0.5$\\

\enddata
\tablenotetext{a}{from \protect\citet{Spolyar et al. b}}
\tablenotetext{b}{in units of g cm$^{-1}$ $s^{-2}$}
\tablenotetext{c}{This observability limit is set by the requirement that a single dark star should be sufficiently bright in at least one JWST filter to give a $5\sigma$ detection after a 100 h exposure, if a gravitational magnification of $\mu=160$ is assumed (see Sect.~\ref{detectability}).}
\end{deluxetable*}

\section{Models for dark stars}
\label{models}

\subsection{Stellar structure and evolution}
As described in the previous section, two physical processes exist for bringing dark matter into a star: gravitational contraction, and capture by nuclear scattering.  Studies of the structure, formation and evolution of dark stars have so far employed one of two simulation techniques: either a `formationary' or a `hydrostatic' approach.  In general, either physical process (or both) can be included with either simulation technique.  Most studies to date employing the formationary approach have included only gravitational contraction, whereas most studies following the hydrostatic approach have only included capture by nuclear scattering.  There are however notable examples of both hydrostatic \citep{Iocco et al.} and formationary \citep{Spolyar et al. b} studies which include both physical processes.

The formationary approach follows the initial collapse of the pre-stellar gas cloud, the resultant gravitational contraction of its dark matter halo, and the subsequent formation of the dark star \citep{Spolyar et al. a,Freese et al. a,Natarajan et al.,Ripamonti et al. a,Spolyar et al. b,Ripamonti et al. b}. This approach captures the salient points of dark star formation and early evolution, which are primarily governed by the annihilation of the gravitationally-contracted dark matter.  The strategy is not optimised for dealing with the stellar evolution after formation, because it relies on either full hydrodynamic simulations \citep{Natarajan et al.,Ripamonti et al. a,Ripamonti et al. b} or analytical approximations to them \citep{Spolyar et al. a, Spolyar et al. b}, becoming either too numerically-demanding or reliant upon simple polytropes once the star has condensed.

The hydrostatic approach assumes some initial model for either a main-sequence \citep{Scott et al. a, Yoon et al., Taoso et al., Scott et al. b, Casanellas & Lopes} or pre-main sequence star \citep{Iocco et al., Umeda et al., Casanellas & Lopes}.  This model is then run through a modified quasi-hydrostatic stellar evolution code \citep[e.g.][]{Scott et al. c} and evolved with the inclusion of energy injection by WIMP annihilation. Late-stage evolution is typically dominated by the dark matter distribution outside the star and the resultant rate at which WIMPs are captured by the nuclear scattering process. Progress has been made in including some early-stage effects, like gravitationally-contracted dark matter \citep{Iocco et al.} and gas accretion \citep{Umeda et al.}, but the realism of such simulations ultimately suffers from their inability to deal with times before hydrostatic equilibrium is reached.  For our purposes, the most significant finding in hydrostatic studies was the possibility that the lifetimes of dark stars might be extended.  In this situation, WIMP-dominated models `stall' (either temporarily or permanently) when they reach an equilibrium configuration somewhere on the Hayashi track \citep{Fairbairn et al., Scott et al. a, Iocco et al., Yoon et al., Taoso et al.}.  For a given stellar mass, the position is essentially dependent only on the WIMP capture rate \citep{Scott et al. b}, and does not depend strongly on whether simulations are started from the main or pre-main sequence \citep{Casanellas & Lopes}.

The formationary approach produces more realistic protostellar structures, whereas the hydrostatic one allows more accurate modelling of long-term evolution.  It is likely that the results of more sophisticated simulations, where formation and later quasi-hydrostatic evolution are treated self-consistently and detailed capture calculations are also included, would resemble a superimposition of the hydrostatic results upon the formationary ones.  Depending upon the time required for dark matter captured by nuclear scattering to become a significant contributor to a star's energy budget, three outcomes are possible:
\begin{enumerate}
\item\label{a}The star may stall directly on the evolutionary paths described by \citet{Spolyar et al. b} during its march toward the main sequence.
\item\label{b}The star may contract onto the ZAMS as per the \citet{Spolyar et al. b} paths, and then re-inflate as annihilation of captured dark matter pushes it back up the Hayashi track.
\item\label{c}The star might travel only some of the way to the ZAMS, but then be turned around and partially re-inflate as the captured dark matter asserts itself.
\end{enumerate}
Scenario \ref{a} would typically be associated with a very quick onset of annihilation from dark matter captured by nuclear scattering, \ref{b} would result if capture takes a very long time to assert itself, and \ref{c} is an intermediate scenario.  The amount of time the captured population takes to become significant is most sensitive to the total capture rate, but also to the time required for equilibrium to be achieved between capture and annihilation, and the time required for WIMPs to thermalise inside the star.  These quantities in turn depend very strongly on the adopted models for the WIMP particle (mass, annihilation cross-section and scattering cross-section) and the dark matter halo (density and velocity structure).  In particular, these dependencies are rather degenerate; the impact of a short equilibrium timescale can for example be mimicked by a denser dark matter halo.  This picture is further complicated by the fact that accretion may also continue to some degree in scenarios \ref{a} and \ref{c}, because it may not have been halted by radiative feedback when the star draws close to the main sequence. The more massive a star becomes, the more dark matter is needed to keep it from contracting onto the main sequence.

Where in the HR diagram a stalled configuration occurs depends upon the total capture rate of dark matter; different annihilation rates are required to support an equilibrium structure at different locations on the Hayashi track.  For a fixed stellar mass, greater rates of capture are required to support a star further from the ZAMS, i.e. at earlier stages in its contraction.  However, larger stellar masses require substantially higher capture rates, so in realistic formation scenarios this effect will be countered to some degree by accretion.  Which effect dominates depends upon the actual accretion rate during the formation phase.

How long the stalling phase persists is determined by the timescale over which the star's hydrogen remains undepleted at a given position on the Hayashi track, and how long capture can be realistically maintained at the given rate. The hydrogen burning timescale is a function of the central temperature and density of the star (i.e. the position on the Hayashi track), so is therefore also a function of the capture rate.  At low capture rates, nuclear burning takes over relatively quickly regardless of how long capture continues, whereas at high capture rates it can in principle remain suppressed (or even entirely absent) for an indefinite period, unless capture drops because the WIMP halo has been depleted.

For very low capture rates -- where stars stall only very briefly near the main sequence (e.g.~very late on the \citealt{Spolyar et al. b} tracks) -- the longest possible lifetime of a dark star is set by the hydrogen-burning lifetime.  This is the approximate maximum time $t_\mathrm{burn}$ that it would take for all the core hydrogen to be converted to helium, if the star were not to contract any further.  For normal Pop III stars, this is set by the timescale in which hydrogen will be depleted by CNO cycle burning, as helium burning by the
triple alpha process quickly creates sufficient C, N and O to catalyze the cycle and allow it to outstrip hydrogen burning by the pp-chain.  In the case of dark stars however, depending upon how close to the ZAMS any
stalling phase occurs, the triple alpha process may not become relevant until rather late.  In this case, the hydrogen burning lifetime will be given by the rate-limiting step of the pp-chain ($p + p \to d + e^+ + \nu_e$), such that 
\begin{equation}
t_\mathrm{burn} = \frac{1}{2m_pr_{pp}(T_\mathrm{c},\rho_\mathrm{c})}.
\end{equation}
Here $m_p$ is the proton mass, $T_\mathrm{c}$ and $\rho_\mathrm{c}$ are the central temperature and density of the star, and $r_{pp}$ is the nuclear reaction rate, given in \citet{Allen's}. For dark stars stalled close
to the ZAMS, the true hydrogen burning lifetime will likely be somewhat shorter, due to the effectiveness of the CNO cycle.  The simulations of \citet{Spolyar et al. b} did not, however, take into account evolution in the stellar
chemical composition, so estimating the time-varying rate of CNO-cycle hydrogen burning in the stars we consider is not completely straightforward. Here, we choose to apply  only the pp-chain limit, as this is a hard limit in the sense that it cannot be evaded by arguments about the stellar model-dependence of any CNO abundances we might derive. 

For very high capture rates, the longest plausible lifetime of a dark star is instead set by the time required for the surrounding dark matter density to drop below that required to sustain the star, due to self-annihilation in the halo.  Once the density of the halo is lowered, capture is reduced and the star contracts and gradually moves onto the main sequence.  The halo density required to power a dark star of a given mass and luminosity can be estimated using Eq.~(14) in \citet{Iocco et al.}.  The self-annihilation time $t_\mathrm{SA}$ for a halo with a given dark matter density can then be obtained by e.g.~inverting Eq.~(6) in \citet{SS09}.  The self-annihilation times obtained in this manner are rather approximate, relying on a number of assumptions we already know to be violated in the first stars (a constant halo self-annihilation rate and the constant infall of WIMPs from a uniform, unbound dark matter halo with an assumed velocity structure).  A more conservative approach, which is thus also much less dependent on the adopted halo model, is to take the corresponding upper limit on stellar lifetimes to be $\sim$$100\, t_\mathrm{SA}$, which is the strategy adopted in this paper. The factor of 100 results from an approximate order of magnitude uncertainty from the constant annihilation rate assumption, and the roughly order of magnitude change seen by \citet{Scott et al. b} in capture rates from WIMP halos with alternative velocity distributions.

The capture and evolutionary histories of the first stars are clearly very dependent upon the adopted dark matter particle and halo models.  To parameterise these uncertainties, as a simple approximation we consider models stalling at various positions on the evolutionary tracks of \citet{Spolyar et al. b}. These models occupy similar positions in the HR diagram to those computed from ZAMS starting models \citep[e.g.][]{Iocco et al.}, despite the rather different physical interpretation of the two scenarios.  We allow stars to stall at various positions on these tracks for times of up to $t_\mathrm{stall}=5\times10^8$\,yr.  This upper limit roughly corresponds to the highest lifetimes considered realistic by \citet{Iocco c}. 
To also take into account the effects of nuclear burning and halo self-annihilation, we limit the plausible lifetimes for any given model to values below
\begin{equation}
t_\mathrm{max} = \mathrm{min}(t_\mathrm{stall},t_\mathrm{burn},100 \, t_\mathrm{SA}).
\label{tmax}
\end{equation}
The parameters for these models are listed in Table~\ref{modeltable}\footnote{Due to convergence problems with the stellar atmosphere models for certain combinations of parameter values, the 100 GeV WIMP track in Table~\protect\ref{modeltable} has one data points less than the track in \protect\citet{Spolyar et al. b}. This does not have any impact on the results from the present paper, since the parameter space sampling of the converged models is more than sufficient to reveal the overall trends in magnitude and colour evolution. The 10 TeV WIMP track also contains one data point more than the Spolyar et al. track to better bracket the 8000--10000 K divide.}.  This age span allows dark stars forming as early as $z = 30$ to survive until $z \approx 10$, and those forming as late as $z = 10$ to survive until the end of reionization ($z \approx 6$). In $5\times 10^8$ yr, the minihalos hosting dark stars are likely to experience many mergers with other minihalos \citep[typically once every $\sim 10^7$ yr][]{Greif et al.}. However, it is not clear whether this prevents or facilitates the existence of long-lived dark stars. Mergers may cause the dark star to be ejected from the center of the halo, thereby cutting off the supply of WIMPs, but may also channel fresh dark matter into the halo center. In principle, one could even consider dark stars with lifetimes in excess of $5\times 10^8$ yr \citep[e.g.][]{Freese et al. c}, perhaps even dark stars surviving into the present-day era, but since the JWST is probably not the optimal telescope for constraining such models, we will not consider them any further in this paper.

Here, we consider models computed with three different WIMP masses (1\,GeV, 100\,GeV and 10\,TeV) in the `minimal capture' approximation of \citet{Spolyar et al. b} in which WIMP annihilation and nuclear fusion contribute equally to the total stellar luminosity; further details can be found in that paper. 

\subsection{Model atmosphere spectra}
To compute the expected spectral energy distributions (SEDs) of dark stars, we have used the MARCS stellar atmosphere code \citep{Gustafsson et al.} for $T_\mathrm{eff}\leq 8000$ K objects and the TLUSTY code \citep{Hubeny & Lanz} for $T_\mathrm{eff}\geq 10 000$ K. Since neither code is able to handle objects in the $T_\mathrm{eff}\approx 8000$--10000 K range, we have interpolated the evolutionary tracks to produce replacement points just outside this temperature region whenever possible. In a few cases, it turned out to be necessary to adopt surface gravities $\log (g)$ slightly different from those given by the \citet{Spolyar et al. b} track to get convergence from MARCS. Neither the omission of the $T_\mathrm{eff}\approx 8000$-10000 K data points, nor the $\log(g)$ deviations have any significant impact on the results of this paper.  

Provided that the properties of dark matter allow the formation of dark stars, objects of this type are expected to be among the first stars forming in the history of the Universe, and therefore to have extremely low metallicities $Z$, possibly at the level given by Big Bang nucleosynthesis \citep[this implies $Z\sim 10^{-9}$, mainly due to Li; e.g.][]{Iocco a}. All models therefore assume primordial abundances of H and He. For computational reasons, the MARCS atmospheres assume an overall metallicity of $Z=2.5\times 10^{-7}$ (corresponding to $\left[ \mathrm{Fe} / \mathrm{H} \right] = -5$ with the $\alpha$-enhanced abundance ratios discussed in \citealt{Gustafsson et al.}), whereas the TLUSTY have been computed at $Z=0$. Tests indicate that this minor inconsistency amounts to an uncertainty in the final JWST magnitudes of $\sim 0.01$ mag, which is irrelevant for the present study.

The TLUSTY models cover the restframe wavelength range 0.015--300 $\mu$m, whereas the MARCS model cover 0.13--20 $\mu$m. While parts of the spectra at wavelengths shortward of 0.13 $\mu$m may be redshifted into the range of the JWST detectors (0.6--29 $\mu$m) at $z>4$, the fluxes in this part of the spectra are too small for detection in the case of cool dark stars. The wavelength coverage of our MARCS models is therefore more than adequate for our needs. 

In the case of the hotter TLUSTY dark stars, corrections for foreground absorption need to be applied at wavelengths shortward of Ly$\alpha$ (0.1216$\mu$m). At these wavelengths, the intergalactic medium becomes increasingly non-transparent at high redshifts due to HI absorption \citep{Madau}, eventually turning almost completely opaque at $z>6$ \citep[e.g.][]{Fan et al.}. To simulate this, all TLUSTY fluxes at $\lambda<0.1216\ \mu$m are set to zero whenever $z>6$, but left unattenuated at lower redshifts. While this treatment may be too crude at lower redshifts \citep{Madau}, this is not a major concern for the current study, which focuses on the prospects of detecting dark stars at $z>6$.

As dark stars evolve along the \citet{Spolyar et al. b} tracks, they eventually attain temperatures similar to those of conventional population III stars ($\sim 10^5$ K) and photoionize large volumes of gas in their surroundings. The resulting HII regions will contribute emission lines and a nebular continuum to the overall spectra of such stars \citep{Schaerer a,Schaerer b}, but this is not taken into account by our models. Because of this, the fluxes that we predict for the very hottest dark stars should be considered conservative.

To obtain JWST magnitudes, all model spectra have been convolved with the transmission profiles for the broadband filters available for the NIRCam (0.6--5 $\mu$m) and MIRI (5--29 $\mu$m) instruments, and calibrated using the AB system. This calibration, which will be used throughout this paper, is based on physical fluxes and defined so that an object with a constant flux per unit frequency interval of 3631 Jy has zero AB-magnitudes $m_\mathrm{AB}$ in all filters.

The rest-frame stellar atmosphere spectra as well as the AB magnitudes (as a function of redshift from $z=0$ to $z=20$ in steps of $\Delta z=0.5$) for the dark star models of Table~\ref{modeltable} are available in electronic format from: http://www.astro.su.se/$\sim$ez

\section{The detectability of high-redshift dark stars}
\begin{figure}
\plotone{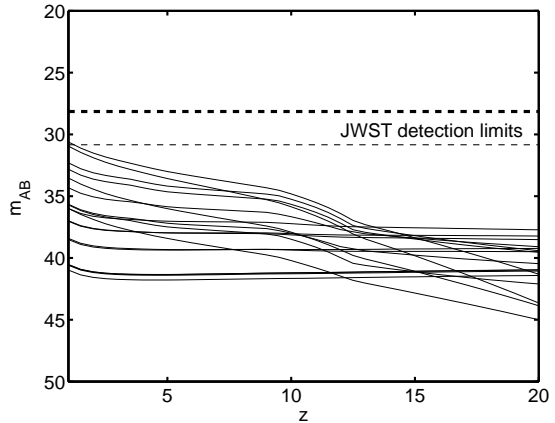}
\caption{The predicted apparent AB magnitudes of dark stars at $z=1$--20 in the NIRCam/F444W filter. Each solid line corresponds to a separate dark star model from Table \ref{modeltable}. The dashed horizontal lines correspond to the JWST detection limits for a $10\sigma$ detection of a point source after $10^4$ s of exposure (thick dashed) and for a $5\sigma$ detection of a point source after $3.6\times 10^5$ s (100 h) of exposure (thin dashed). At $z=10$--20, the dark stars are 4--14 magnitudes too faint for detection. Hence, without the magnification boost of a foreground galaxy cluster, JWST will not be able to detect individual population III dark stars at the redshifts where they formed. Long-lived dark stars surviving until the end of reionization ($z\approx 6$) appear somewhat brighter, but are still at least 2 magnitudes below the detection limit. 
\label{ABmag_nolens}}
\end{figure}

\label{detectability}
\begin{figure*}
\plottwo{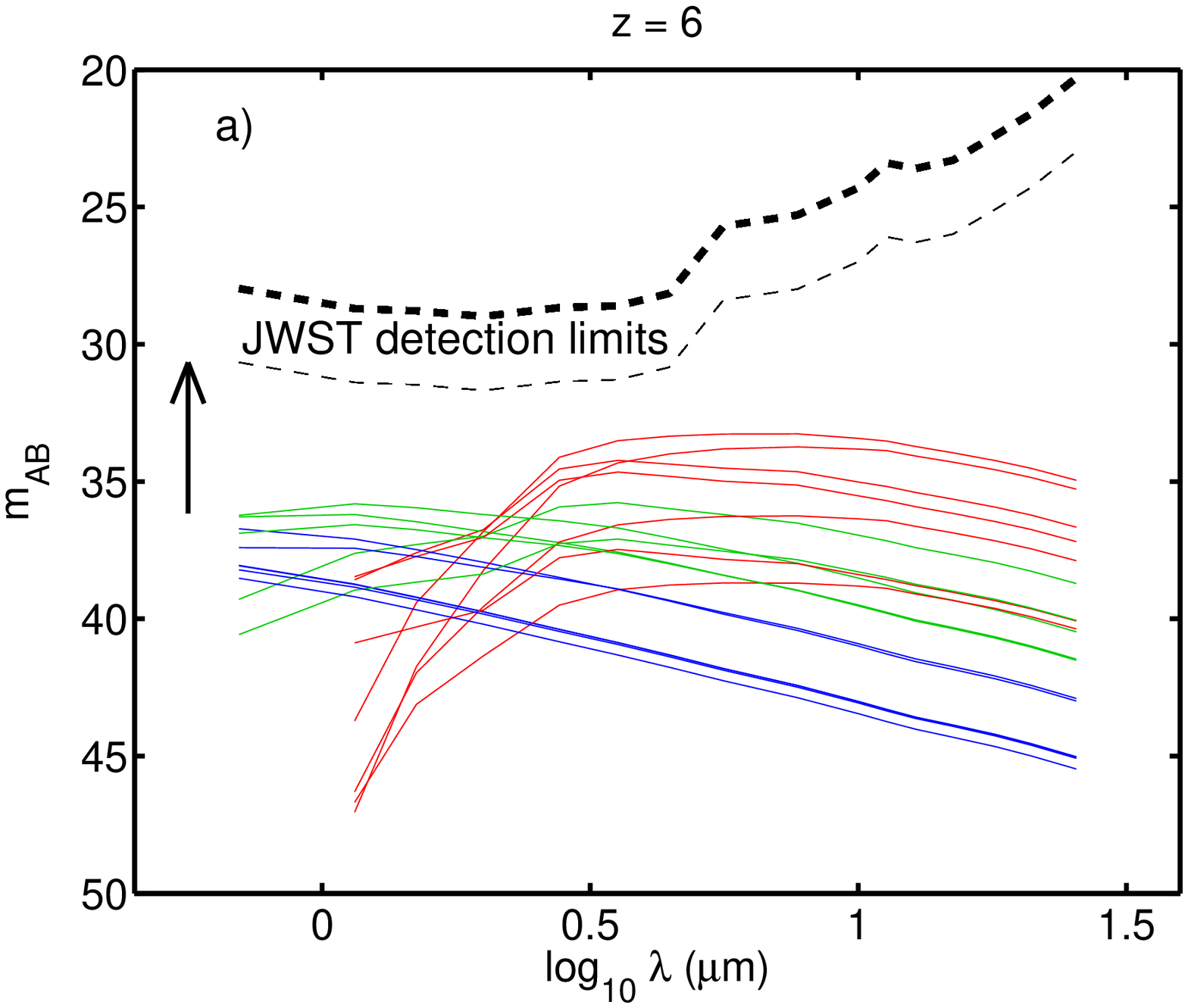}{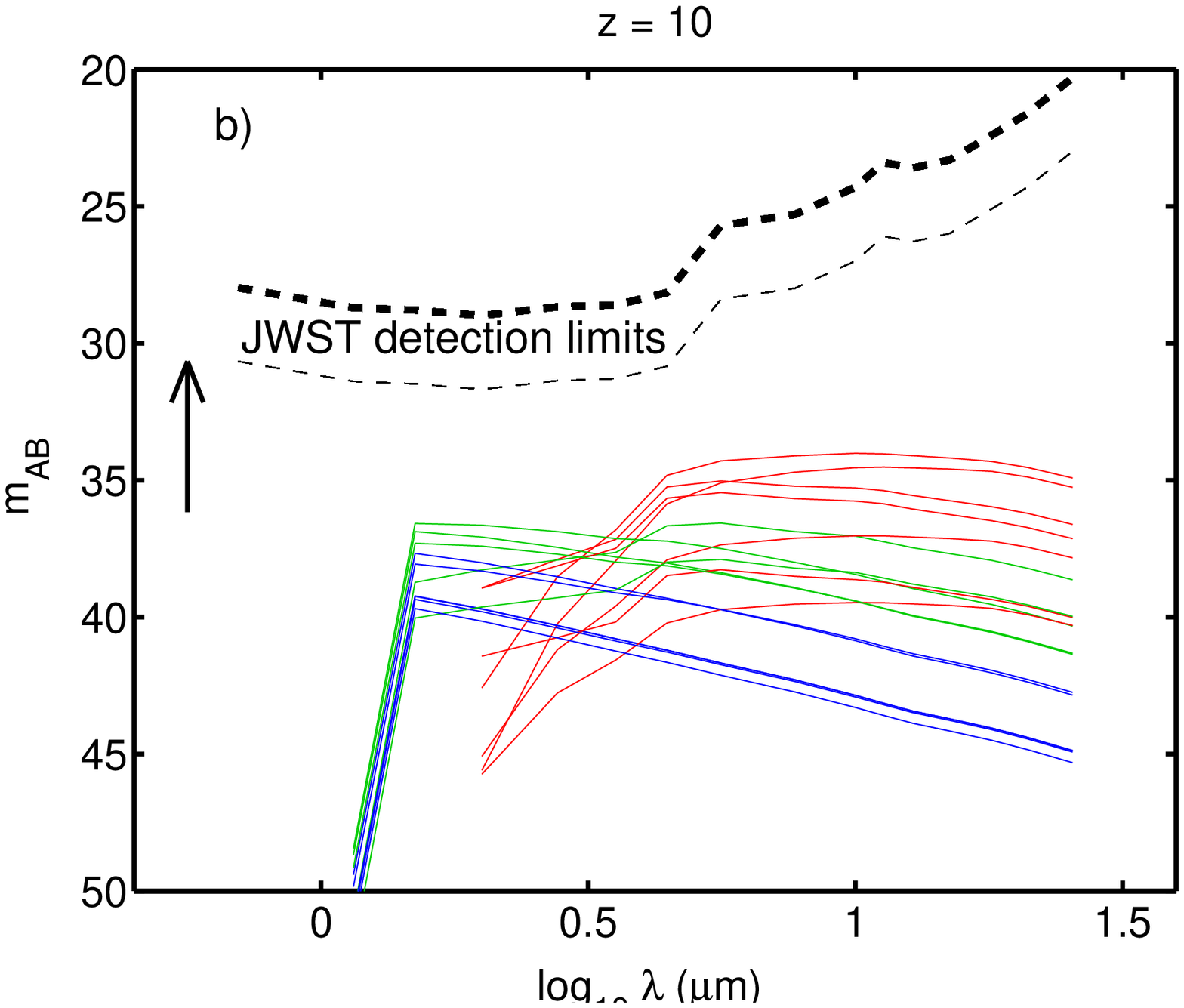}
\caption{The predicted apparent AB magnitudes of dark stars at $z=6$ ({\bf a}) and $z=10$ ({\bf b}), as a function of central wavelength of the JWST broadband filters. Each solid line corresponds to a separate dark star model from Table \ref{modeltable}. These lines have been colour-coded according to the effective temperatures of the dark stars: $T_\mathrm{eff}\leq 8000$ K (red), $8000\ \mathrm{K} < T_\mathrm{eff} \leq 30000\ \mathrm{K}$ (green) and $T_\mathrm{eff}>30000$ K (blue). The dashed horizontal lines correspond to the JWST detection limits for a $10\sigma$ detection of a point source after $10^4$ s of exposure (thick dashed) and for a $5\sigma$ detection of a point source after $3.6\times 10^5$ s (100 h) of exposure (thin dashed). The progressively brighter detection thresholds at central wavelengths higher than 4.4$\mu$ ($\log_{10}\lambda>0.65$) reflect the lower sensitivity of the MIRI instrument (central filter wavelengths $\log_{10}\lambda>0.65$) compared to NIRCam (central filter wavelengths $\log_{10}\lambda\leq 0.65$). In both panels, all dark star models lie significantly faintward of the detection thresholds in all filters, implying that their intrinsic brightnesses are too low to be detected by JWST. However, a magnification of $\mu=160$ due to gravitational lensing by a foreground galaxy cluster (see Sect.~\ref{lensing}) would shift all models upward by 5.5 magnitudes (as indicated by the vertical arrow) and allow certain varieties of dark stars into the brightness regime detectable by the NIRCam instrument. This is the case for some of the $T_\mathrm{eff}\leq 30000$ K dark stars (green and red lines) at both $z=6$ and $z=10$. The reason why the red lines end abruptly at 1.15 $\mu$m ($\log_{10} \lambda=0.06$) for $z=6$ and at 2.0 $\mu$m ($\log_{10} \lambda=0.3$) for $z=10$ is that the short-wavelength limit (0.13 $\mu$m) of the MARCS model spectra have entered the bluer filters at these redshifts. Since this happens at $m_\mathrm{AB}>38$, which is a brightness regime inaccessible to the JWST, this has no impact on the present study. The sharp drop in brightness at
$\lambda \leq 1.5\ \mu$m ($\log_{10} \lambda\leq 0.17$) along the blue and green lines at $z=10$ is due to HI absorption in the foreground intergalactic medium.
\label{ABmag_nolens2}}
\end{figure*}

\subsection{Dark stars in random fields}
In Fig.~\ref{ABmag_nolens}, we present the AB magnitudes of all dark star models from Table \ref{modeltable} as a function of redshift ($z=1$--20) in the NIRCam F444W filter. Also included are two estimated JWST detection thresholds for point sources, indicated by dashed horizontal lines. The first one is based on $10\sigma$ detections for $10^4$ s exposures (thick dashed) and the second on $5\sigma$ detections after 100 h ($3.6\times10^5$ s) exposures (thin dashed). The former represent the fiducial JWST detection limits listed on the JWST homepage\footnote{http://www.jwst.nasa.gov/}, whereas the latter roughly correspond to the magnitude limits expected for `ultra deep field' type observations.
\begin{figure*}
\plottwo{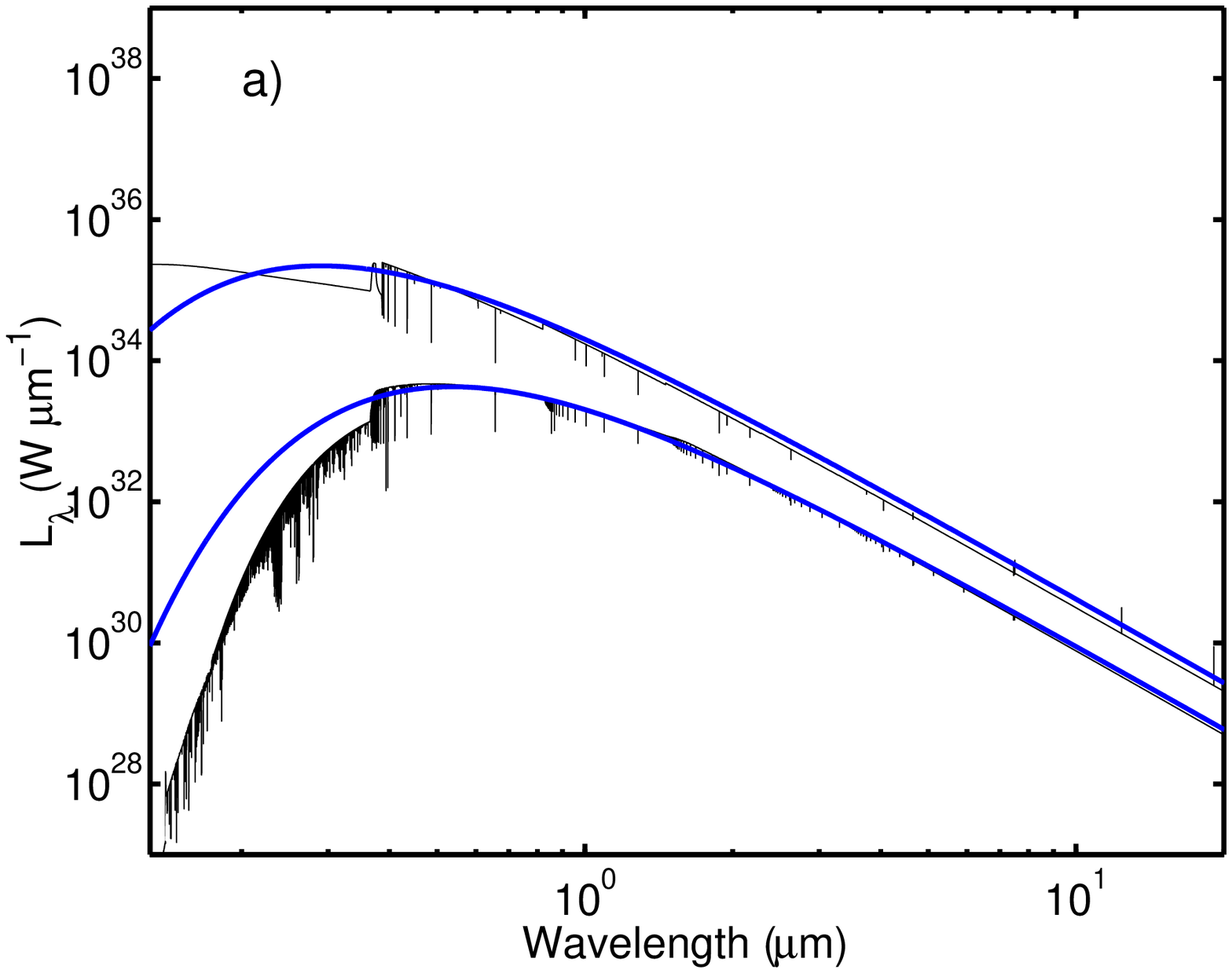}{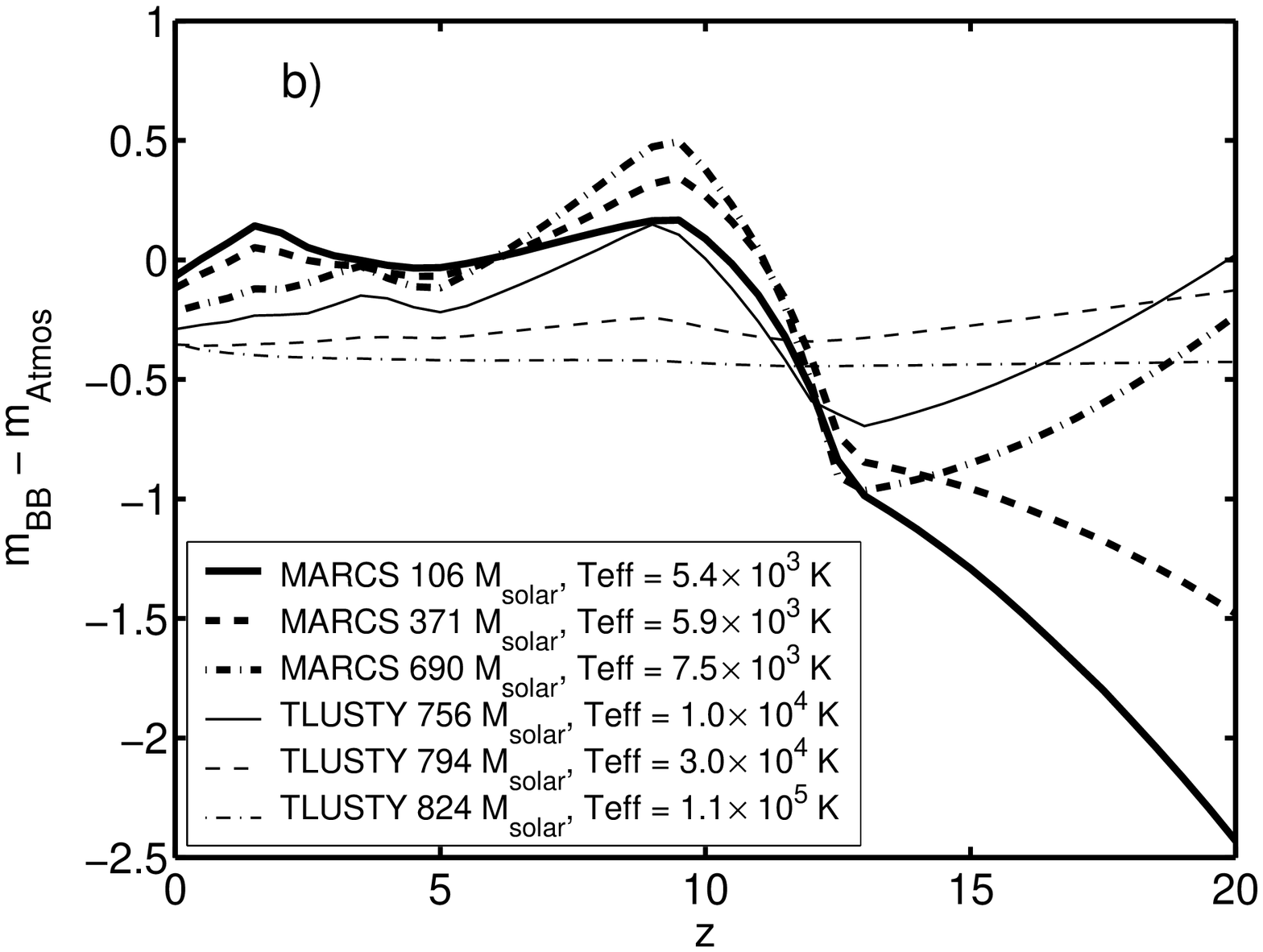}
\caption{Synthetic stellar atmosphere spectra compared to black body spectra for dark stars. {\bf a)} The SEDs of the 106 $M_\odot$, $T_\mathrm{eff}=5400$ K (lower SEDs) and the 756 $M_\odot$, $T_\mathrm{eff}=10000$ K (upper SEDs, multiplied by 100 to avoid cluttering) dark stars predicted for a 1 GeV WIMP. The black lines correspond to the synthetic stellar atmosphere SEDs generated by MARCS (for the lower, $T_\mathrm{eff}=5400$ K spectra) and TLUSTY (for the upper $T_\mathrm{eff}=10000$ K spectra), whereas the blue lines correspond to black body spectra generated for identical temperatures and bolometric luminosities. The obvious differences are the lack of breaks (most notable at $\approx 0.36\ \mu$m) and absorption lines in the black body spectra. Due to the low but non-zero metallicity ($\left[ \mathrm{Fe} / \mathrm{H} \right] = -5$) of the MARCS spectrum for the 106 $M_\odot$, $T_\mathrm{eff}=5400$ K dark star, numerous metal lines are seen shortward of $0.36\ \mu$m in the lower SED. While these lines appear to be strong due to the high spectral resolution of the model spectrum, they actually have have very small equivalent widths and negligible impact on the JWST broadband fluxes. 
{\bf b)} The difference between the black body AB magnitudes $m_\mathrm{BB}$ and the synthetic stellar atmosphere AB magnitudes $m_\mathrm{Atmos}$ in the NIRCam F444W filter. The different lines represent the six dark star models from Table~\ref{modeltable} for a 1 GeV WIMP. The solid lines correspond to the MARCS (thick line) and TLUSTY (thin lines) dark star SEDs plotted in {\bf a)}. There are substantial differences between $m_\mathrm{BB}$ and $m_\mathrm{Atmos}$ for the cooler ($\lesssim 10000$ K) dark stars, whereas the differences for hotter dark stars are below 0.5 mag. This is primarily because the hotter stars become progressively more black body-like at the relevant wavelengths. For instance, the 0.36 $\mu$m break (which gives rise to the negative $m_\mathrm{BB}-m_\mathrm{Atmos}$ at $z\gtrsim 9$) is far less prominent for hotter objects.
\label{Atmos_vs_bb}}
\end{figure*}

It is immediately clear that all dark star models lie significantly below both these detection thresholds at high redshifts. At $z=10$--20, the intrinsic luminosities of the dark stars convert into apparent magnitudes that are 4--14 magnitudes too faint. Hence, without the magnification boost of a foreground galaxy cluster, JWST will not be able to detect individual population III dark stars at the redshifts where they formed. Long-lived dark stars surviving until the end of reionization ($z\approx 6$) appear somewhat brighter, but are still at least 2 magnitudes below the detection limit. 

Some of the dark star models in Fig.~\ref{ABmag_nolens} exhibit F444W magnitudes which change very little as a function of redshifts. 
This is due to the steep the spectra of the $T_\mathrm{eff}>20000$ K dark stars, which attain their peak fluxes at rest wavelengths $<0.2\mu$m (i.e. on the short-wavelength side of the F444W filter for all redshifts in the plotted range). As the redshift is increased, intrinsically brighter parts of their spectra are redshifted into the F444W filter, and this almost exactly compensated for the increased luminosity distance.

In Fig.~\ref{ABmag_nolens2}, we display AB magnitudes for all dark stars from Table~\ref{modeltable} at $z=6$ and $z=10$ as a function of the central wavelengths of all broad JWST filters. The JWST detection limits, defined as in Fig.~\ref{ABmag_nolens}, are also included. All dark star models from Table~\ref{modeltable} lie significantly faintward of these thresholds, regardless of which filter is used. However, a magnification of $\mu=160$ due to gravitational lensing by a foreground galaxy cluster (see Sect.~\ref{lensing}) would shift all models upward by $\approx 5.5$ magnitudes (as indicated by the vertical arrow) and shift certain varieties of dark stars into the brightness regime detectable by the NIRCam instrument (but not by MIRI). This is the case for several of the $T_\mathrm{eff}\leq 8000$ K dark stars (red lines) and a couple of the $8000 <T_\mathrm{eff}\leq 30000$ K dark stars (green lines) at both $z=6$ and $z=10$. 

In Fig.~\ref{Atmos_vs_bb} we demonstrate the need for using synthetic stellar atmosphere spectra (as compared to pure black body spectra) for predictions of this type. In Fig.~\ref{Atmos_vs_bb}a, the MARCS spectra for the 106 $M_\odot$ ($T_\mathrm{eff}=5400$ K) dark star and the TLUSTY spectra for the 756 $M_\odot$ ($T_\mathrm{eff}=10000$ K) dark star from the 1 GeV WIMP track in  Table~\ref{modeltable} are compared to black body spectra based on identical temperatures and bolometric luminosities. At rest frame wavelengths of $\lambda>0.4\ \mu$m, the shape of the continua are very similar for the stellar atmosphere and black body SEDs, but the presence  of breaks in the stellar atmosphere spectra (most notably the Balmer break at $\lambda\approx 0.36\ \mu$m) will introduce substantial differences once these are redshifted into the JWST filters. The stellar atmosphere SEDs also contain a large number of absorption lines, but these will have far smaller effect of the broadband fluxes discussed here. In Fig.~\ref{Atmos_vs_bb}b, we show the difference between the magnitudes derived from black body and the synthetic stellar atmosphere spectra in the NIRCam F444W filter. The different lines represent the six dark star models from Table~\ref{modeltable} for a WIMP mass of 1 GeV. The solid lines correspond to the MARCS (thick lines) and TLUSTY (thin lines) dark star SEDs plotted in Fig.~\ref{Atmos_vs_bb}a. As seen, there are substantial differences for the cooler ($\lesssim 10000$ K) dark stars, whereas the differences for hotter dark stars are below 0.5 mag. This is primarily because the hotter stars become progressively more black body-like at the relevant wavelengths. For instance, the $0.36\ \mu$m break (which makes the black body spectra overpredict the F444W fluxes at $z\gtrsim 9$) is far less prominent in the hotter dark stars. We conclude, that while black body spectra may be useful for deriving order-of-magnitude estimates of JWST fluxes for the hotter dark stars, detailed stellar atmosphere models are required to accurately predict the fluxes for cool dark stars at high redshifts.  
 
\subsection{Dark stars magnified by gravitational lensing}
\label{lensing}
Fig.~\ref{ABmag_nolens} and ~\ref{ABmag_nolens2} demonstrate that all dark stars in Table~\ref{modeltable} are {\it intrinsically} too faint at $z\geq 6$ to be detected by JWST, even if extremely long exposure times ($t_\mathrm{exp}=3.6\times 10^5$ s, i.e. 100 h) are considered. The only hope of detecting isolated dark stars with JWST at these redshifts would then be to exploit the gravitational lensing provided by a foreground galaxy cluster. Galaxy clusters at $z\approx 0.1$--0.6 can in principle boost the fluxes of high-redshift objects by up to factors $\sim 100$ \citep[e.g.][]{Bradac et al.,Maizy et al.}. As shown in Fig.~\ref{ABmag_nolens2}, this would be sufficient to lift some of the cooler ($T_\mathrm{eff}<30000$ K) dark star models above the JWST detection threshold. For each separate dark star model, the entries in the $\max(z_\mathrm{obs})$ column in Table~\ref{modeltable} indicate the maximum redshifts at which $5\sigma$ detections are possible in at least one JWST filter after $3.6\times 10^5$ s (100 h) exposures, assuming a magnification of $\mu=160$ (see below). No dark stars are detectable at $z>13$, even when this boost due to lensing is taken into account. 

How many dark stars at $z\approx 10$ would one then expect to detect in a survey of a single lensing cluster? This depends on the magnification properties of the cluster, on the cosmic star formation history of dark stars and on their typical lifetimes $\tau$ (which we treat as a free parameter bounded from above by $t_\mathrm{max}$, as discussed in section \ref{models}: $0\leq \tau \leq t_\mathrm{max}$). While gravitational lensing boosts the fluxes of background objects, their surface number densities are at the same time diluted by a factor equal to the magnification $\mu$. In a region with angular area $\theta^2$ and magnification $\mu$, one can show that the expected number of dark stars $N_\mathrm{DS}$ in the redshift interval $\left[z_\mathrm{min},z_\mathrm{max} \right]$ is given by:
\begin{equation}
N_\mathrm{DS}=c\theta^2 \int_{z_\mathrm{min}}^{z_\mathrm{max}} \int_{t(z)}^{t(z)-\tau} \frac{\mathrm{SFR}(t)d_\mathrm{os}(z)^2 (1+z)^3}{\mu(z) M_\mathrm{DS}}  \frac{\mathrm{d}t}{\mathrm{d}z} \mathrm{d}t \ \mathrm{d}z
\label{N_DS}
\end{equation}
where $\mathrm{SFR}(t)$ is the star formation rate (in units of $M_\odot\ \mathrm{Mpc}^{-3} \mathrm{yr}^{-1}$) of dark stars at cosmic epoch $t(z)$, $M_\mathrm{DS}$ is the dark star mass (assumed to be the same for all such objects) and $d_\mathrm{os}(z)$ is the angular size distance between the observer and source at redshift $z$. In the case of flat $\Lambda$CDM cosmologies, the derivative $\frac{\mathrm{d}t}{\mathrm{d}z}$ in eq.(\ref{N_DS}) is given by:
\begin{equation}
\frac{\mathrm{d}t}{\mathrm{d}z}=-\frac{1}{H_0 (1+z)\left[\Omega_{M}(1+z)^3 + \Omega_\Lambda \right]^{1/2}}
\label{dt_dz}
\end{equation}

When exploring the prospects of detecting isolated dark stars in the high-magnification regions of a foreground galaxy cluster, we have adopted MACS J0717.5+3745 at $z=0.546$ as our lensing cluster. This object has the largest angular Einstein radius detected so far, with a relatively shallow surface mass-density profile which boosts the projected area corresponding to high magnifications \citep{Zitrin et al. a,Zitrin et al. b}. Both of these properties combine to make this the best lens currently known for the study of faint objects at very high redshifts. For this cluster, the angular area over which the magnification is in the range $\mu=100$--300 (with an average magnification $\overline{\mu}\approx 160$) for sources at $z_\mathrm{s}=6$--20 is $\approx 0.3$ arcmin$^2$. 

\begin{figure*}
\centering
\plottwo{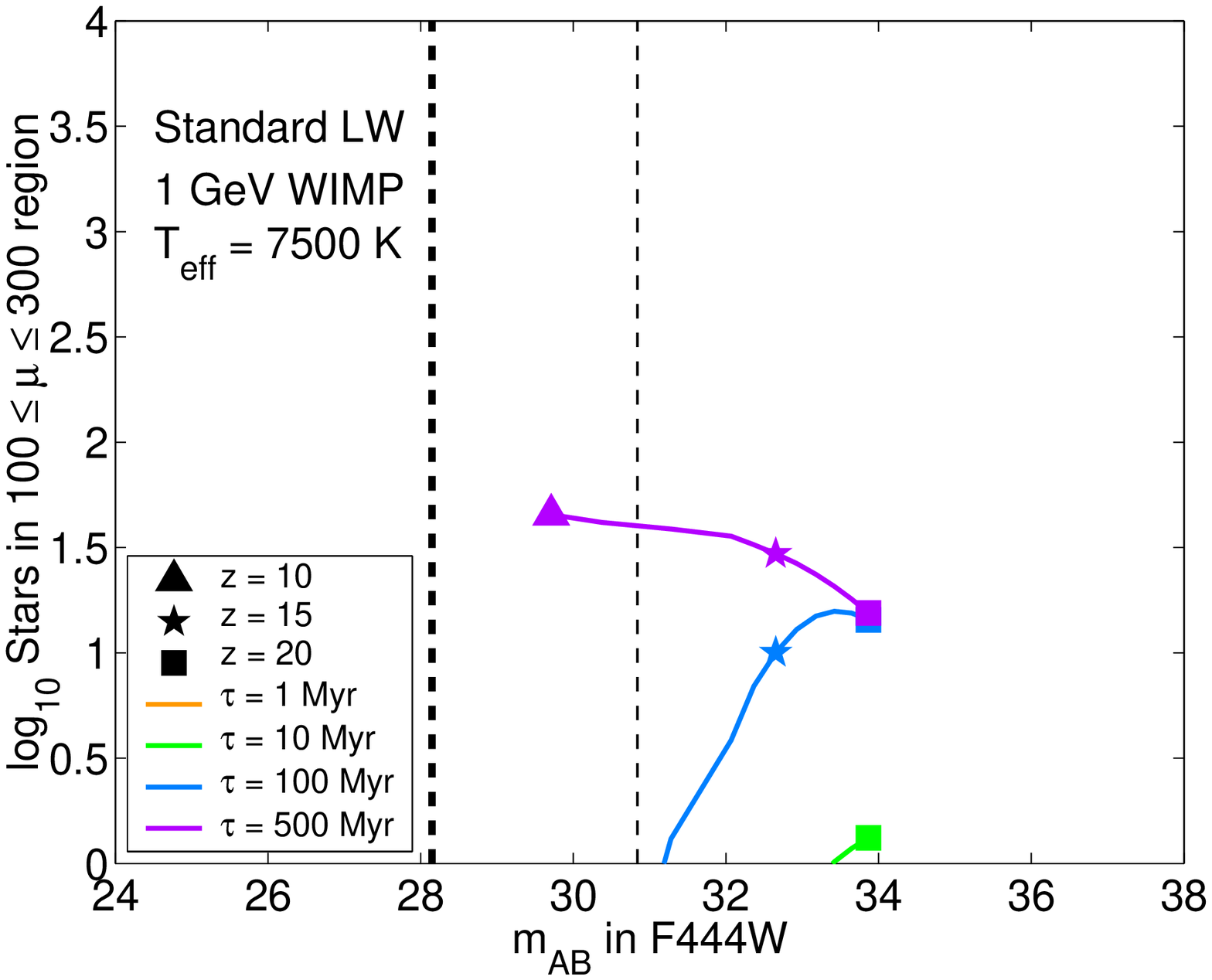}{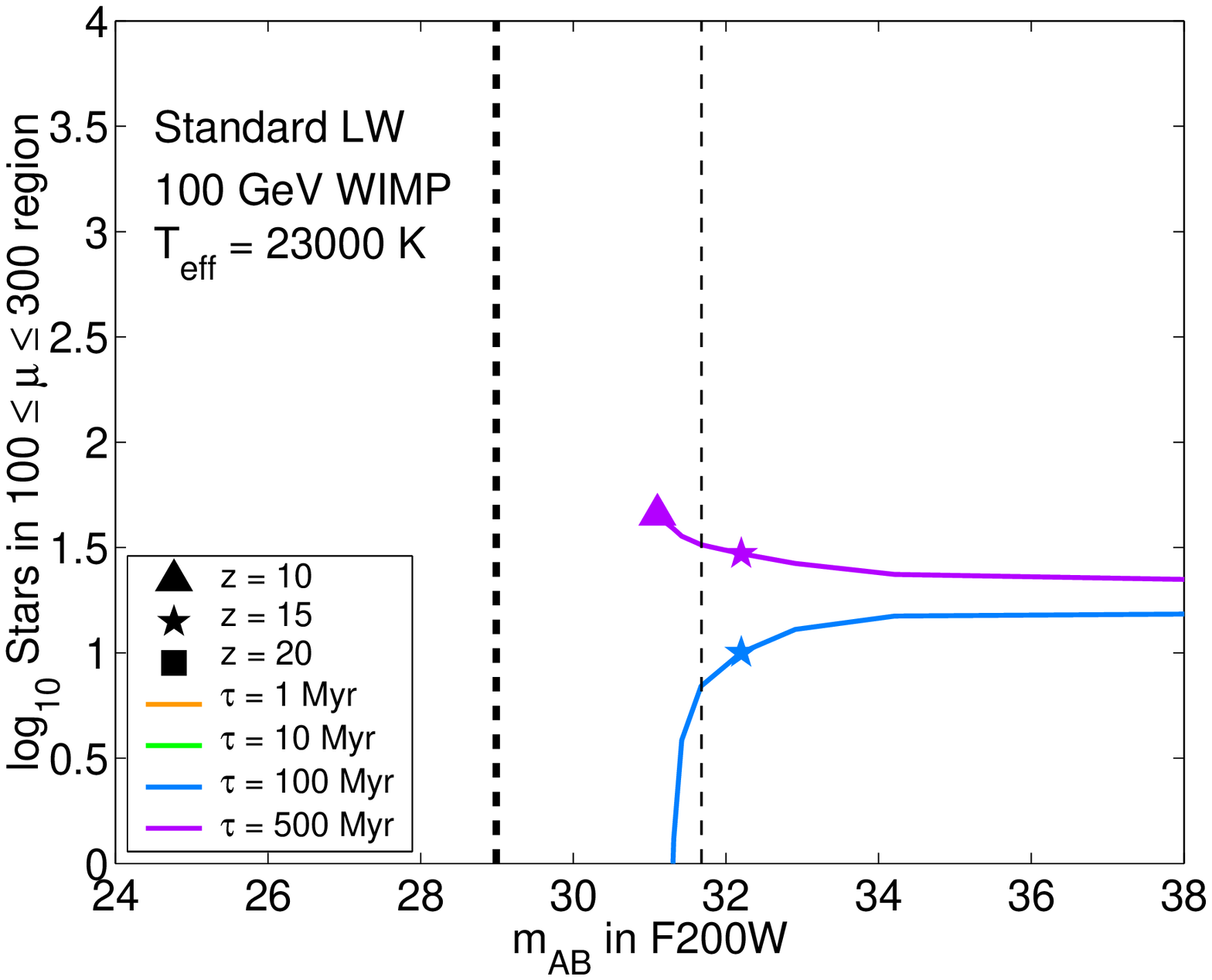}
\plottwo{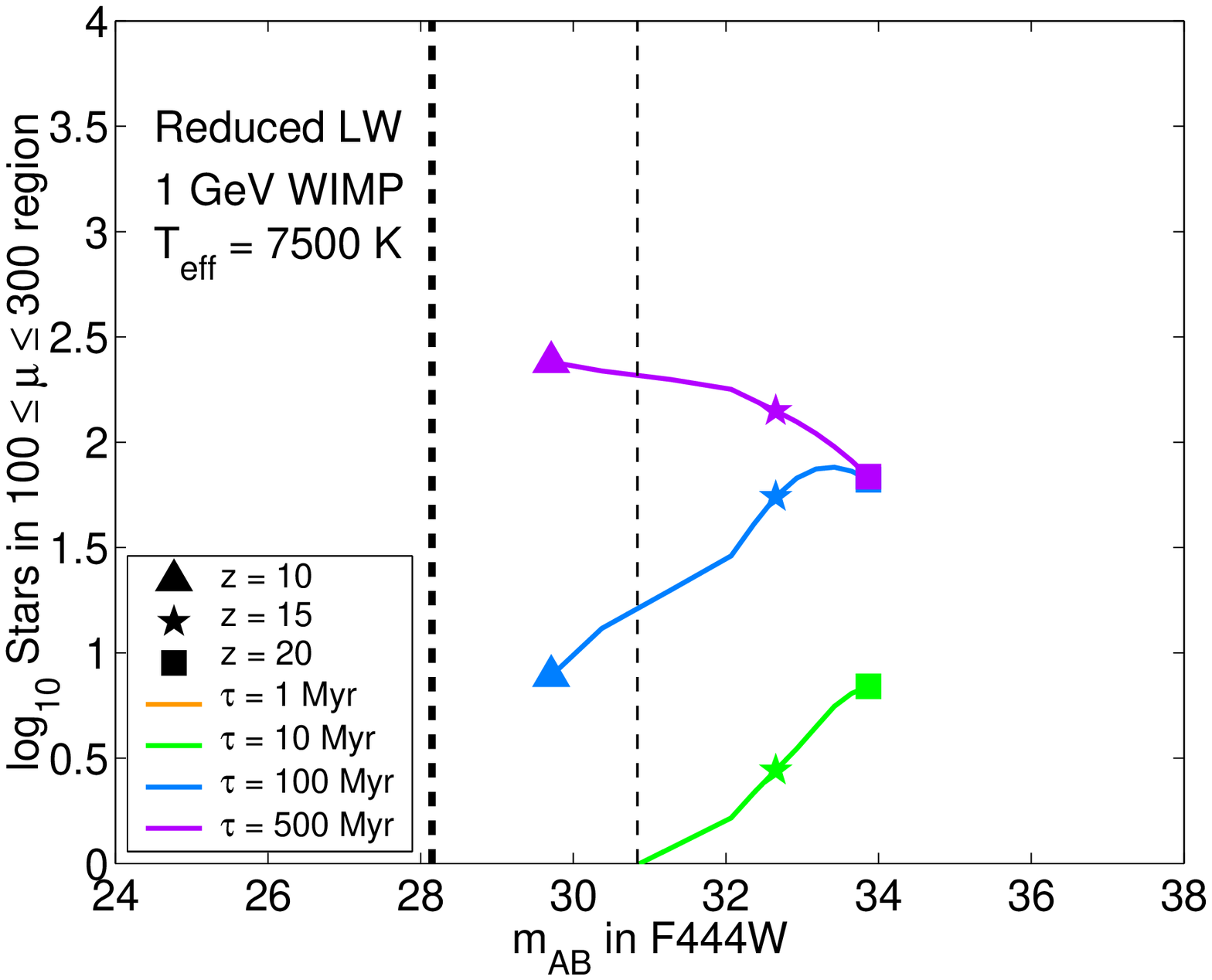}{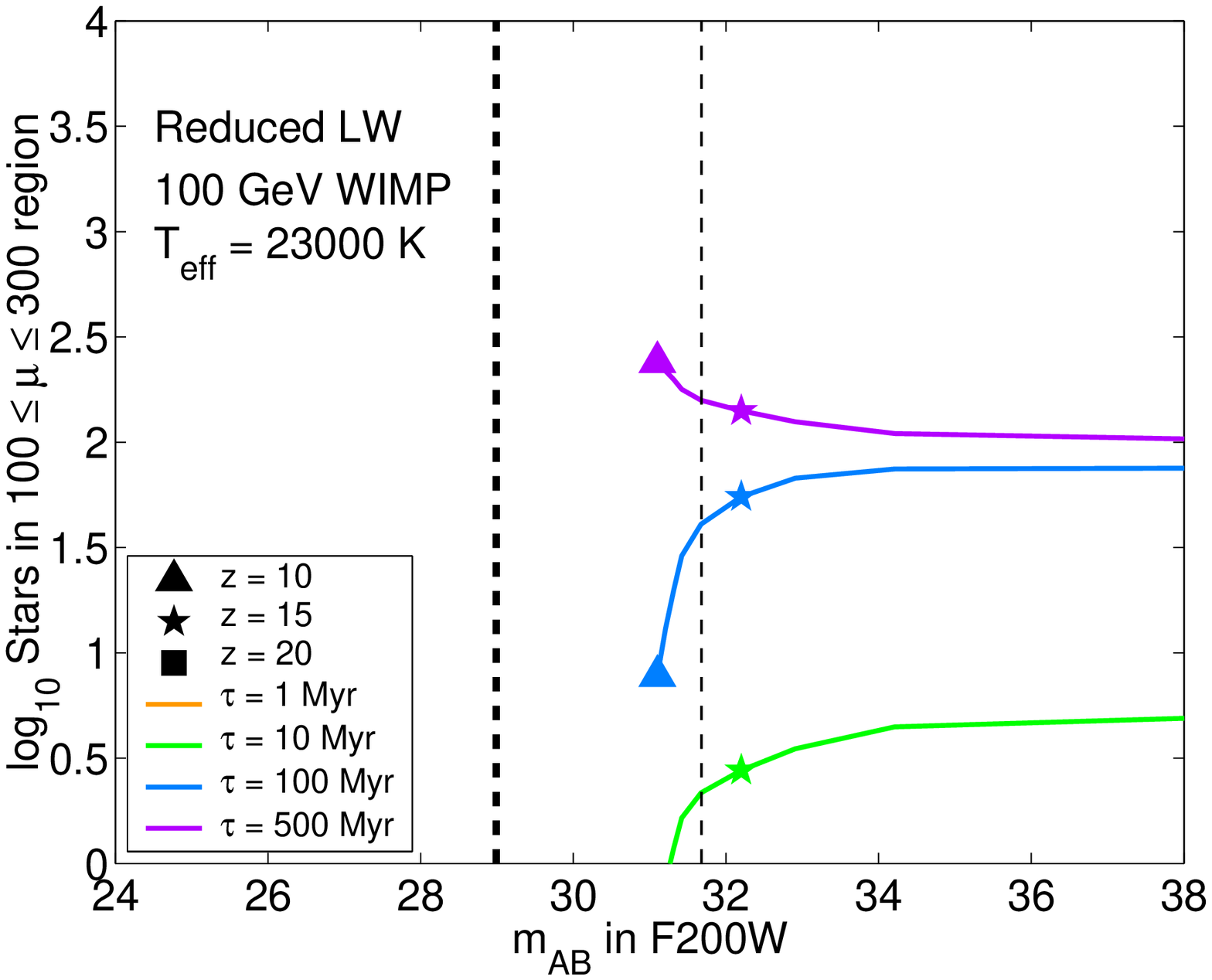}
\plottwo{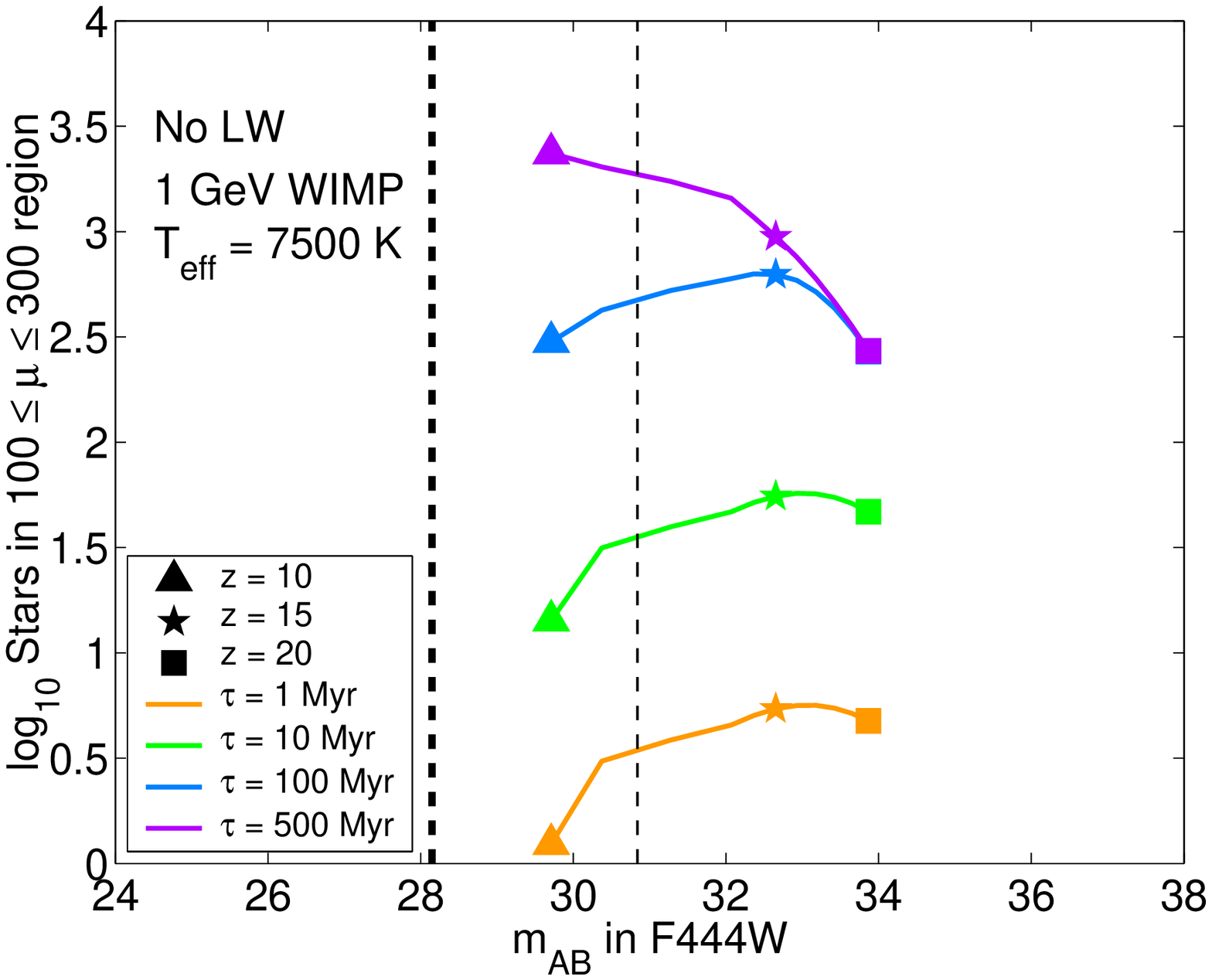}{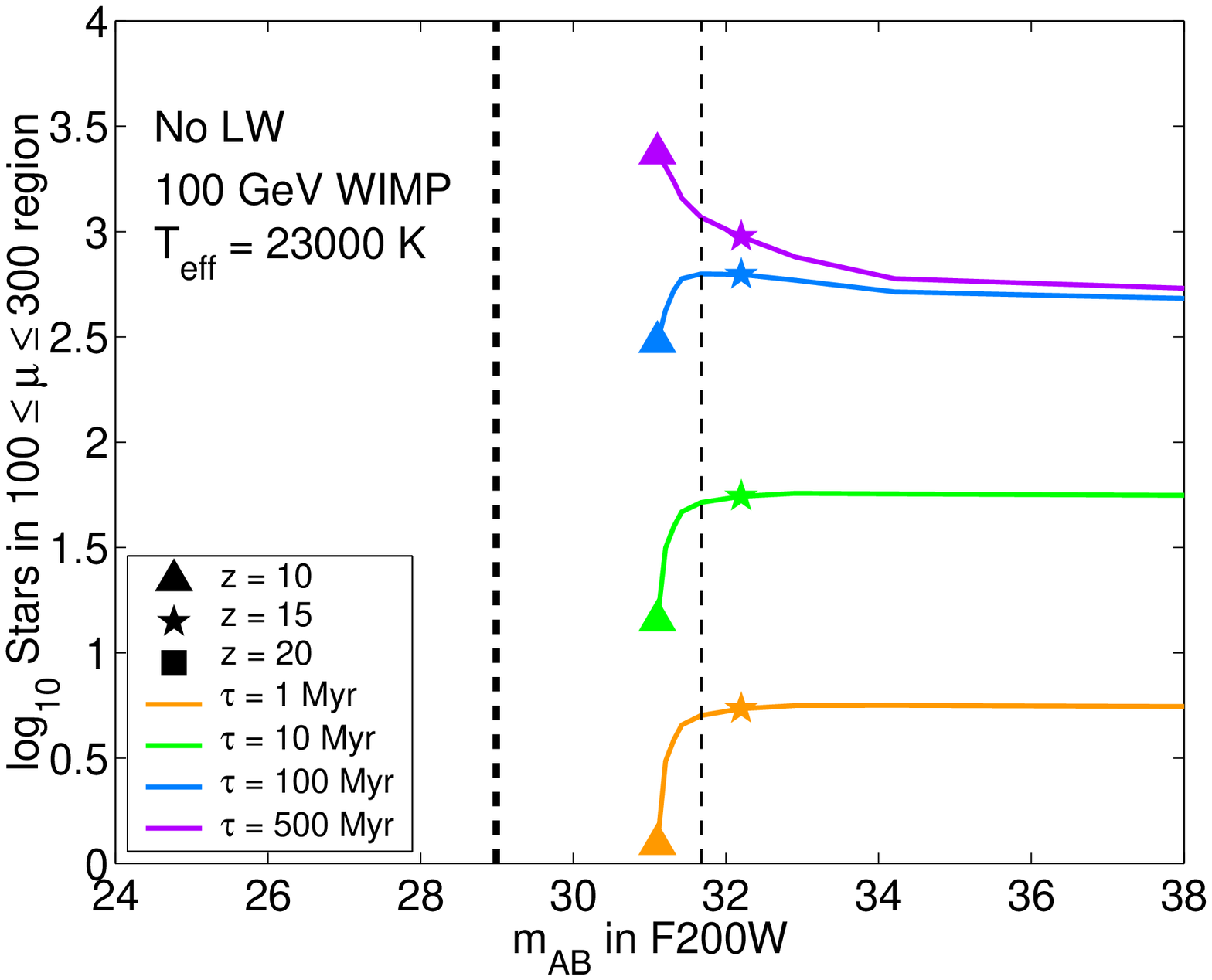}
\caption{The number of dark stars per unit redshift interval predicted within the high-magnification regions ($\mu=100$--300) of the galaxy cluster MACS J0717.5+3745, as a function of their apparent AB magnitudes in the JWST/NIRCam F444W (left column) and F200W (right column) filters. In the left column, the 690 $M_\odot$ dark star model with $T_\mathrm{eff}=7500$ K from the 1 GeV WIMP track has been used, and in the right column the 716 $M_\odot$ dark star with $T_\mathrm{eff}=23000$ K from the 100 GeV WIMP track. Within each panel, the differently coloured lines correspond to dark star life times of $\tau=1$ Myr (orange), 10 Myr (green), 100 Myr (blue) and 500 Myr (purple). Symbols along these lines indicate dark star redshifts of $z=10$ (triangle), $z=15$ (star) and $z=20$ (square).  The different rows correspond to the standard LW (top row), reduced LW (middle row) and no LW (bottom row) star formation histories for population III stars in minihalos \citep{Trenti & Stiavelli}. The vertical lines within each panel represent the JWST detection thresholds for $10\sigma$ detections after $10^4$ s exposures (thick dashed) and $5\sigma$ detections after $3.6\times 10^5$ s (100 h) exposures (thin dashed). All population III stars forming through H$_2$ cooling in minihalos are here assumed to go through a dark star phase with identical properties, but the results can easily be generalized by scaling the curves downward by the dark star fraction $f_\mathrm{DS}$. The lines extending to the right hand borders of the plots reflect the sharply decreasing fluxes in the F200W filter due to foreground HI absorption in the intergalactic medium at $z>15$.  
\label{N_mag}}
\end{figure*}

For the cosmic star formation history $\mathrm{SFR}(t)$ of dark stars, we have explored three different scenarios from \citet{Trenti & Stiavelli} for single population III stars forming through H$_2$ cooling in minihalos. These three scenarios differ in their assumptions concerning the amount of Lyman-Werner (hereafter LW) feedback expected during the relevant epochs. LW radiation is emitted by hot, high-mass stars, destroys H$_2$ and inhibits further population III star formation along this cooling channel. The very massive, fusion-driven population III stars that \citet{Spolyar et al. b} envision as the descendants of dark stars will emit copious amounts of LW radiation. However, if many of the population III stars that form through H$_2$ cooling \citep[population III.1 stars in the notation suggested by][]{Greif & Bromm} go through a long-lived dark star phase (during which LW fluxes are insignificant due to the very low effective temperatures of these objects), this could substantially delay the onset of the LW feedback compared to scenarios with no dark stars. Hence, the amount of LW feedback at a given epoch is expected to depend both on the fraction $f_\mathrm{DS}$ of population III.1 stars that at some point become dark stars, and the typical duration $\tau$ of this phase. 

The three scenarios from \citet{Trenti & Stiavelli} that we consider are hereafter referred to as {\it standard LW}, {\it reduced LW} and {\it no LW}, and correspond to the cosmic star formation histories of population III stars formed through H$_2$ in minihalos depicted in their Figs. 1, 2 and 3, respectively. The standard LW scenario, which we consider suitable for very small $f_\mathrm{DS}$ and short $\tau$, pushes the bulk of population III star formation to higher redshifts than the other two (possibly more suitable for large $f_\mathrm{DS}$ and $\tau$). Whereas all population III star formation has effectively ceased by $z\approx 10$ in the standard LW model, significant formation of such stars continues to lower redshifts in the other two scenarios.

In Fig.~\ref{N_mag}, we plot the apparent AB magnitudes of selected dark star models versus the number of dark stars expected within the high-magnification regions ($\mu=100$--300; $\overline{\mu}\approx 160$) of MACS J0717.5+3745 per unit redshift interval. These plots are based on the assumption of $f_\mathrm{DS}=1$, i.e. that {\it all} population III stars formed through H$_2$ cooling in minihalos become dark stars, and that all such dark stars display similar properties (in terms of mass, temperature, radius and lifetime). In reality, this may not be very realistic. The individual merger history of each minihalo is likely to give rise to some variation in the central CDM density. Some minihalos may also host multiple stars \citep{Turk et al.,Stacy et al.}, which could cause some or all of the population III stars forming in such systems to wander out of the minihalo centre where dark matter annihilation is the most efficient. The predictions in Fig.~\ref{N_mag} can, however, easily be rescaled to other dark star fractions by shifting the curves downward by a factor $f_\mathrm{DS}$. 

The left column of Fig.~\ref{N_mag} displays the results for the 690 $M_\odot$, $T_\mathrm{eff}=7500$ K dark star from the 1 GeV WIMP track and the right column the corresponding results for the 716 $M_\odot$, $T_\mathrm{eff}=23000$ K  dark star from the 100 GeV WIMP track. Due to the different temperatures of these dark stars, the AB-magnitudes in the NIRCam F444W filter are used in the left column and the NIRCam F200W filter in the right column. For both of these dark star models, the upper limit on the lifetime set by eq.~(\ref{tmax}) is $t_\mathrm{max}=5\times 10^8$ yr. For each model, we therefore consider lifetimes of $\tau=10^6$ (orange lines), $10^7$ (green lines), $10^8$ (blue lines) and $\tau=5\times 10^8$ yr (purple lines). The three rows of panels in Fig.~\ref{N_mag} correspond to the cosmic population III star formation histories with standard LW (top row), reduced LW (middle row) and no LW feedback (bottom row). The different markers within each panel indicate the AB-magnitudes and numbers of dark stars at redshifts $z=10$ (triangle), $z=15$ (star) and $z=20$ (square).

In general, long dark star lifetimes $\tau$ imply larger numbers of dark stars at detectable brightnesses. In the case of the 690 $M_\odot$, $T_\mathrm{eff}=7500$ K dark star (right column), considerable numbers ($\geq 10$) of dark stars are expected to be sufficiently bright for detection (at 5$\sigma$ after $3.6\times 10^5$ s) in the high-magnification regions of MACS J0717.5+3745 at $z\approx 10$--11, provided that their lifetimes are $\approx 5\times 10^8$ yr. This holds regardless of which of the three feedback scenarios is adopted. For the reduced LW and no LW feedback models, significant dark stars are expected even if the lifetime is closer to $\tau=10^7$ yr. In the no LW scenario, even $\tau=10^6$ yr dark stars can be detected behind MACS J0717.5+3745, but only in modest numbers ($<10$). The situation for the 716 $M_\odot$, $T_\mathrm{eff}=23000$ K dark star (right column) is similar, except that this model remains detectable up to a redshift of $z\approx 13$. 

Dark stars with $\tau=5\times 10^8$ yr do not show the same decline in their expected numbers when going from $z=15$ to $z=10$ as dark stars with shorter lifespan do. This happens because $\tau=5\times 10^8$ yr dark stars have lifetimes that exceed the cosmic age intervals between adjacent redshift bins, allowing such objects to accumulate at lower redshifts, even though the cosmic star formation rate of population III stars is declining at these epochs for all three feedback scenarios. The \citet{Trenti & Stiavelli} models do not allow us to trace the star formation history to epochs at $z<10$, even though the star formation rate clearly remains non-zero at $z=10$ in both the reduced LW and no LW feedback scenarios. In fact, $\tau=5\times 10^8$ yr dark stars forming at $z\geq 10$ in these feedback scenarios will survive in detectable numbers to even lower redshifts ($z\approx 6$), even if one artificially sets their formation rates to zero at $z<10$.

In summary, cool ($T_\mathrm{eff}\leq 30 000$ K) and long-lived ($\tau \gtrsim 10^7$ yr) dark stars may well be detected at $z\approx 10$ in sizeable numbers within a single, ultra deep JWST field if one takes advantage of the magnifying power of a foreground galaxy cluster. In the case of high $\tau$ ($\gtrsim 10^8$ yr) and cosmic star formation scenarios which imply significant dark star formation at $z<15$, several dark stars may be seen even if only a minor fraction ($f_\mathrm{DS}\sim 0.01$--0.1) of all population III stars forming in minihalos become dark stars with temperatures in the detectable range.

\section{Discussion}
\label{discussion}
\begin{figure}
\plotone{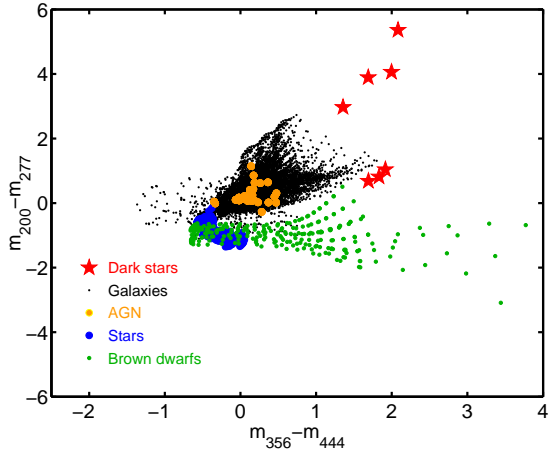}
\caption{The JWST/NIRCam $m_{356}-m_{444}$ vs. $m_{200}-m_{277}$ colours of $T_\mathrm{eff}<10000$ K dark stars at $z=10$ (red star symbols) compared to a number of potential interlopers in multiband surveys: star clusters or galaxies at $z=0$--15 (black dots), AGN template spectra at $z=0$--15 (yellow dots), Milky Way stars with $T_\mathrm{eff}=2000$-50000 K and $Z=0.001-0.020$ (blue dots) and Milky Way brown dwarfs with $T_\mathrm{eff}=130$--2200 K (green dots). Since the dark stars reside a region of this colour-colour diagram that is disconnected from those occupied by these other objects, it should be possible to identify possible dark star candidates in deep multiband JWST/NIRCam surveys.  
\label{colour_criteria}}
\end{figure}

\subsection{How to distinguish isolated dark stars from other objects}
As demonstrated in Sect.~\ref{lensing}, certain varieties of $z\approx 10$ dark stars may be sufficiently bright and numerous to be detected by a JWST/NIRCam survey of the high-magnification regions of a foreground galaxy cluster. But how does one identify such objects among the overwhelming number of mundane interlopers located in front of, inside or beyond the lensing cluster? 

Given the many degeneracies involved in the interpretation of broadband photometry, there may well be unresolvable ambiguities in some cases. However, many of the cooler high-redshift dark stars should stand out in multiband survey data because of their unusual colours.
This is demonstrated in Fig.~\ref{colour_criteria}, where we plot the colour indices $m_{356}-m_{444}$ vs. $m_{200}-m_{277}$ (based on AB-magnitudes in the JWST/NIRCam F200W, F277W, F356W and F444W filters) at $z=10$ for all dark stars from Table~\ref{modeltable} with $T_\mathrm{eff}< 10000$ K. The colours of these models (red star symbols) are compared to the colours predicted for a wide range of galaxies, star clusters, active galactic nuclei (AGN) and Milky Way stars. The cloud of black dots in Fig.~\ref{colour_criteria} indicate the colours of integrated stellar populations (star clusters and galaxies) generated with the \citet{Zackrisson et al.} spectral synthesis model. These predictions are based on instantaneous-burst\footnote{This is a conservative choice, since allowing for more extended star formation histories would only result in a more restricted colour coverage for these objects}, Salpeter-IMF stellar populations at redshifts $z=0$--15 with metallicities in the range $Z=0.001$-0.020, ages ranging from $10^6$ yr up to the age of the Universe at each redshift and a rest-frame stellar dust reddening of $E(B-V)=0$--0.5 mag assuming the \citet{Calzetti et al.} extinction law. Also included in Fig.~\ref{colour_criteria} are the expected colours of foreground stars with $T_\mathrm{eff}=2000$-50000 K and $Z=0.001-0.020$ in the Milky Way (i.e. at $z=0$), based on the \citet{Lejeune et al.} compilation of synthetic stellar atmosphere spectra (blue dots), and the colours of Milky Way brown dwarfs in the 130--2200 K range based on the \citet{Burrows et al. a,Burrows et al. b} models (green dots). The yellow dots represent the template AGN spectra of \citet{Hopkins et al.} for bolometric luminosities $\log_{10} L_\mathrm{bol}/L_\odot=8.5$--14.0, at redshifts $z=0$--15. Since none of these potential interlopers have $m_{356}-m_{444}$ and $m_{200}-m_{277}$ colours that overlap with those of cool $z\approx 10$ dark stars, a diagnostic diagram of this type can be used to cull objects that are clearly {\it not} dark stars from multiband survey data. However, given that $T_\mathrm{eff}< 10000$ K dark stars are unlikely to attain apparent magnitudes brighter than $m_{AB}\approx 30$ at their peak wavelengths (even when boosted by gravitational lensing; see Fig.~\ref{ABmag_nolens2}b), and can realistically only be detected in one or two NIRCam filters, follow-up spectroscopy of the remaining dark star candidates will be required to establish their exact nature. This will admittedly be very challenging, but a very coarse spectrum can possibly be obtained for $m_{AB}\approx 30$ objects with JWST/NIRSpec (assuming a $3.6\times 10^5$ s exposure). Objects of this type may also be suitable targets for the 42 m  European Extremely Large Telescope\footnote{http://www.eso.org/sci/facilities/eelt/} (E-ELT).

\subsection{The spectral signature of dark stars in high-redshift galaxies}
\begin{figure}
\plotone{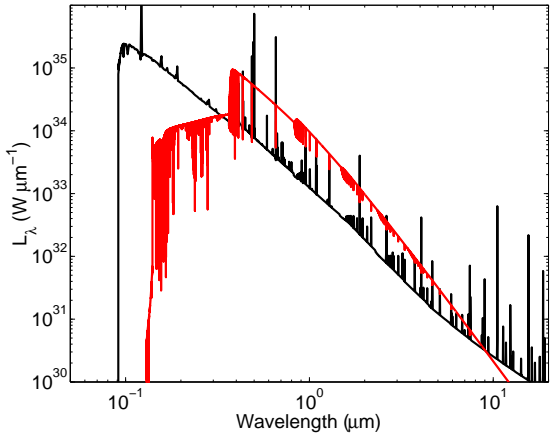}
\caption{The rest-frame spectrum of a $10^6\ M_\odot$, $Z=0.001$, Salpeter-IMF stellar population (stellar mass range 0.08--120 $M_\odot$) which has formed stars at a constant rate for $10^8$ yr (black line) with a superimposed contribution (red line) from ten $T_\mathrm{eff}=7500$ K, 690 $M_\odot$ dark stars (from the 1 GeV WIMP track). Despite making up only $0.7\%$ of the stellar mass in this galaxy, these dark stars give rise to an upturn in the spectrum at rest-frame wavelengths longward of 0.36$\mu$m. In a $z=10$ object, this spectral feature should appear longward of 3.96$\mu$m and could serve as a telltale signature of cool dark stars within the first galaxies. Due to the low but non-zero metallicity ($\left[ \mathrm{Fe} / \mathrm{H} \right] = -5$) of the MARCS model for the 690 $M_\odot$ dark stars, numerous metal absorption lines are seen throughout the red spectrum. While these lines appear to be strong due to the high spectral resolution of the model, they actually have very small equivalent widths and negligible impact on the JWST broadband fluxes.
\label{firstgal}}
\end{figure}

The first galaxies are expected to form at $z\approx 10$--13 in CDM halos with total masses around $10^8\ M_\odot$ \citep[e.g.][]{Johnson et al. a,Johnson et al. b,Greif et al.,Ricotti}. At the time of assembly, each such object is likely to contain a number of minihalos in which population III stars have already formed \citep{Greif et al.}. If some of these population III stars go through a long-lived ($\tau\gtrsim 10^8$ yr) dark star phase, several dark stars may in principle congregate inside the first generation of galaxies and give rise to telltale signatures in their integrated spectra. In Fig.~\ref{firstgal}, we display the rest-frame spectrum predicted by the \citet{Zackrisson et al.} population synthesis model for a $10^8$ yr old, low-metallicity ($Z=0.001$, i.e. population II), Salpeter-IMF (mass range 0.08--120 $M_\odot$) stellar population which has formed stars at a constant rate. This population has been assigned a stellar mass of $10^6 \ M_\odot$, which -- for a $10^8\ M_\odot$ halo with baryon fraction $f_\mathrm{bar}\approx 0.7$ -- corresponds to $\sim 10\%$ of the baryonic mass in stars, in rough agreement with the models of \citet{Ricotti} for a $z\approx 10$ galaxy. The predicted NIRCam magnitudes of this object lie around $m_{AB}\approx 33$--34 which would make it sufficiently bright for detection with JWST if seen through a gravitational lens with magnification $\mu\gtrsim 10$. Superposed on the spectrum of this high-redshift galaxy is the integrated contribution from ten $T_\mathrm{eff}=7500$ K (the 690 $M_\odot$ models from the 1 GeV WIMP track) dark stars (red line). These dark stars, which contribute only $0.7\%$ of the stellar mass in this galaxy, give rise to a conspicuous red bump in the spectrum at rest-frame wavelengths longward of 0.36$\mu$m (this corresponds to wavelengths longer than 3.96$\mu$m at $z=10$). Because of this, galaxies that contain many cool dark stars are expected to display anomalously red colours. A feature like this is very difficult to produce through other means. For instance, the spectrum depicted in Fig.~\ref{firstgal} cannot be attributed to dust reddening, since no known extinction law would allow a sharp rise in flux at wavelengths longward of 0.36$\mu$m to co-exist with a very blue continuum at shorter wavelengths. Attributing the red bump as due to thermal dust emission is also untenable, since this would require dust radiating at a temperature close to that of the dark star (7500 K), which is higher than the sublimation temperature of all known types of dust. 

\begin{figure}
\plotone{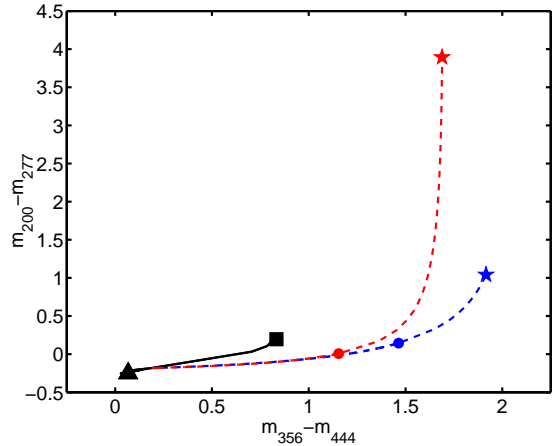}
\caption{The JWST/NIRCam $m_{356}-m_{444}$ vs. $m_{200}-m_{277}$ colour evolution as a function of age for a  $z=10$, low-metallicity ($Z=0.001$), Salpeter-IMF galaxy experiencing a short burst of star formation ($10^8$ yr) and passive evolution thereafter (black line). The black triangle indicates an age of $10^6$ yr and the black square $5\times 10^8$ yr (roughly the age of the Universe at this redshift). The star symbols indicate the colours of two cool, $z=10$ dark stars from Table~\ref{modeltable}: the 7500 K, $690 \ M_\odot$ model from the 1 GeV WIMP track (blue star) and the 5800 K, $106 M_\odot$ model from the 100 GeV WIMP track (red star). Both of these (as is the case for all $T_\mathrm{eff}<10000$ K dark stars observed in these filters; see Fig.~\ref{colour_criteria}) are considerably redder than the colours expected for galaxies, regardless of their age. Dashed lines indicate how the colours of the model galaxy (at an assumed age of $10^8$ yr) would shift if it were to contain dark stars of either of the two types. The filled circles along the dashed tracks indicate mixtures at which these dark stars make up $1\%$ of the stellar mass in the model galaxy. These points are also significantly redder than the reddest point along the galaxy track, indicating that a $\sim 1\%$ stellar mass fraction in dark stars within $z\approx 10$ galaxies would be detectable through multiband photometry.
\label{colour_criteria2}}
\end{figure}

In Fig.~\ref{colour_criteria2}, we display the $m_{356}-m_{444}$ vs. $m_{200}-m_{277}$ colour evolution (black solid line) as a function of age for the synthetic galaxy from Fig.~\ref{firstgal} at $z=10$. Here, the age runs from $10^6$ yr (black triangle) up to the age of the Universe at this redshift ($\approx 5\times 10^8$ yr). Also indicated are the colours of two cool dark stars: the 7500 K model from the 1 GeV WIMP track (blue star) and the 5800 K model from the 100 GeV WIMP track (red star). Due to their low temperatures, these two dark stars are far redder than the model galaxy in both colours plotted. As shown in Fig.~\ref{colour_criteria}, this is generally the case for $T_\mathrm{eff}<10000$ K dark stars observed in these filters. The galaxy is here assumed to go through a short burst of star formation (forming stars at a constant rate for $10^8$ yr), after which it evolves passively. This gives a conservative estimate of the colour difference between galaxies and dark stars. Allowing a more extended star formation episode or a star formation rate that increases over time would only increase the discrepancy between the colours of dark stars and the model galaxy. The dashed lines indicate how the colours of a $10^8$ yr galaxy would shift in this diagram, if it were to harbour dark stars of the type considered. The filled circles along each such mixing track indicate the position at which the dark stars make up $1\%$ of the total stellar mass. Since these points are significantly redder in the $m_{356}-m_{444}$ colour than the reddest point along the standard galaxy track, a $\sim 1\%$ stellar mass fraction in dark stars would result in very peculiar colours for $z=10$ galaxies and should allow such objects to be identified as candidate `dark star galaxies' in JWST multiband survey data.

\section{Summary}
\label{summary}
In this paper, we have investigated the prospects of detecting very massive, high-redshift dark stars using the JWST. While individual dark stars at $z>6$  will be intrinsically too faint for detection, we demonstrate that the magnification provided by a foreground galaxy cluster will make certain varieties of long-lived ($\tau\geq 10^7$ yr) and cool ($T_\mathrm{eff}\leq 30000$ K) detectable at redshifts up to $z\approx 10$. We argue that it should be possible to identify at least some of these dark stars in photometric NIRCam surveys due to their peculiar colours. If the lifetimes of dark stars are sufficiently long, they may also congregate during the hierarchical assembly of the first galaxies. We find that this could give rise to distinct signatures in the integrated NIRCam colours of high-redshift galaxies, provided that dark stars make up at least $\sim 1\%$ of the overall stellar mass in these objects.

\acknowledgments
EZ, CER and G\"O acknowledge grants from the Swedish National Space Board. EZ, PS, CER, BE, G\"O and SS acknowledge funding support from the Swedish Research Council. FI acknowledges support from the 7$^{th}$ European Community research program
FP7/2007/2013 within the framework of convention \#235878. G\"O is a Royal  Swedish Academy of Sciences Research Fellow, supported by a grant from the Knut an Alice Wallenberg foundation. PG acknowledges partial support by NSF Award PHY-0456825. The authors are indepted to Marcia J. Rieke and Kay Justtanont for giving us access to the NIRCam and MIRI broadband filter profiles prior to public release. We also thank the anonymous referee, who provided very useful comments on the manuscript.

\end{document}